\documentclass[aps,prd,onecolumn,notitlepage,superscriptaddress,nofootinbib]{revtex4-1}  
\usepackage{graphicx}  
\usepackage{bm}        
\usepackage{amssymb}   
\usepackage{amsmath}
\usepackage{comment}
\usepackage[hidelinks]{hyperref}
\usepackage{eufrak}
\usepackage{bbm}
\usepackage{color}
\usepackage{subcaption}
\usepackage{import}
\usepackage{dipoles}
\usepackage[subpreambles=true]{standalone}
\usepackage{graphicx}
\usepackage{xfrac}

\begin{document}
\author{Kevin Dusling}
\email{kdusling@mailaps.org}
\affiliation{American Physical Society, 1 Research Road, Ridge, New York 11961, USA}
\affiliation{Physics Department, Brookhaven National Laboratory, Bldg. 510A, Upton, New York  11973, USA}
\author{Mark Mace}
\email{mark.mace@stonybrook.edu}
\affiliation{Physics Department, Brookhaven National Laboratory, Building 510A, Upton, New York  11973, USA}
\affiliation{Physics and Astronomy Department, Stony Brook University, Stony Brook, New York 11974, USA}
\author{Raju Venugopalan}
\email{raju@bnl.gov}
\affiliation{Physics Department, Brookhaven National Laboratory, Bldg. 510A, Upton, New York 11973, USA}

\newcommand{\mfm}[1]{{{\color{blue} #1}}}

\newcommand{\Bp}{B_p}
\newcommand{\lqcd}{\Lambda_{\rm QCD}}
\newcommand{\Qst}{Q_{s,T}}
\newcommand{\Qs}{Q_{s}}
\newcommand{\pmax}{p_\perp^{\rm max}}

\newcommand{\Eq}[1]{Eq.~(\ref{#1})}
\newcommand{\Eqs}[2]{Eqs.~(\ref{#1})-(\ref{#2})}
\newcommand{\Fig}[1]{Fig.~(\ref{#1})}
\newcommand{\Figs}[2]{Fig.~(\ref{#1})-(\ref{#2})}

\title{Parton model description of multiparticle azimuthal correlations in $pA$ collisions}
\date{\today}

\begin{abstract}
In \cite{Dusling:2017dqg}, an initial state ``parton model" of quarks scattering off a dense nuclear target was shown to qualitatively reproduce the systematics of multiparticle azimuthal anisotropy cumulants measured in proton/deuteron-nucleus ($pA$) collisions at RHIC and the LHC. The systematics included
i) the behavior of the four-particle cumulant $c_2\{4\}$, which generates a real four-particle second Fourier harmonic $v_2\{4\}$,
ii) the ordering $v_2\{2\}>v_2\{4\}\approx v_2\{6\}\approx v_2\{8\}$ for two-, four-, six-, and eight-particle Fourier harmonics,
iii) the behavior of so-called symmetric cumulants $\text{SC}(2,3)$ and $\text{SC}(2,4)$.
These features of azimuthal multiparticle cumulants were previously interpreted as a signature of hydrodynamic flow; our results challenge this interpretation. We expand here upon our previous study and present further details and novel results on the saturation scale and transverse momentum ($p_\perp$) dependence of multiparticle azimuthal correlations. We find that the dependence of $v_2\{2\}$ and $v_2\{4\}$ on the number of color domains in the target varies with the $p_\perp$ window explored. We extend our prior discussion of symmetric cumulants and compute as yet unmeasured symmetric cumulants. We investigate the $N_c$ dependence of $v_2\{2\}$ and $v_2\{4\}$.  We contrast our results, which include multiple scatterings of each quark off the target, to the Glasma graph approximation, where each quark suffers at most two gluon exchanges with the target. We find that coherent multiple scattering is essential to obtain a positive definite $v_2\{4\}$. We provide an algorithm to compute expectation values of arbitrary products of the ``dipole" lightlike Wilson line correlators.
\end{abstract}

\maketitle

\section{Introduction}
In Ref.~\cite{Dusling:2017dqg}, we demonstrated that a simple initial state parton model gives rise to many of the features of multiparticle azimuthal correlations observed experimentally in small collision systems at both RHIC at BNL and the LHC at CERN. In this model, collinear quarks from the projectile scatter coherently off color sources of the size of the inverse of the saturation scale $Q_s$ in the nuclear target; the scattered quarks are found to be collimated in their relative azimuthal angles. We observed crucially that one obtains a negative value for $c_2\{4\}$, the four-particle azimuthal anisotropy cumulant. This results in a positive definite four-particle Fourier coefficient $v_2\{4\}$. We demonstrated further, in a simpler Abelian version of our model, that one obtains the ordering of $m$-particle second Fourier harmonics $v_2\{2\}>v_2\{4\}\approx v_2\{6\}\approx v_2\{8\}$. Both of these features were previously believed to be unique signatures of collectivity arising from the hydrodynamic flow of quark-gluon matter. Not least, we demonstrated that  so-called symmetric cumulants (mixed four-particle cumulants of different Fourier harmonics) computed in this parton model display the same qualitative features as experimental measurements of symmetric cumulants. We note that symmetric cumulants were designed to probe correlations and fluctuations arising from the hydrodynamic response to different harmonics of the azimuthal structure of the initial geometry.

While the hydrodynamic description of the flow of quark-gluon matter is likely valid in the larger, more central, collisions of heavy-ions ($AA$), the applicability of this description to more peripheral $AA$ collisions, and to $pA$ and proton-proton ($pp$) collisions, is less clear~\cite{Bzdak:2013zma,Niemi:2014wta}.  Our results in Ref.~\cite{Dusling:2017dqg} are therefore a strong hint that the stated measures of hydrodynamic collectivity are not robust in their own right without further corroboration from other distinct measures of collectivity. An example of the latter is the strong jet quenching that is seen in central heavy-ion collisions at RHIC and the LHC~\cite{Adams:2003im,Adams:2003kv,Abelev:2012hxa,CMS:2012aa}. In contrast, jet quenching is either small or absent in peripheral $AA$ collisions and in $pA$ collisions~\cite{Adams:2003im,Adam:2016dau,McGlinchey:2017esf}.

We will elaborate in this paper on the parton model description introduced in Ref.~\cite{Dusling:2017dqg}, which extends previous work on two-particle azimuthal correlations discussed in \cite{Lappi:2015vha,Lappi:2015vta}. A novel feature is the development of a general algorithm (based on the framework in \cite{Blaizot:2004wv}) to compute expectation values of multi-dipole correlators.  These objects encode the physics of multiple eikonal scattering of quarks on a colored target.  In particular, the dipole operator is the trace over the product of a lightlike Wilson line appearing in the quark production amplitude at a given transverse position, with its conjugate transpose appearing in the complex-conjugate amplitude at a different spatial location, normalized by the number of colors $N_c$. We will present a systematic study of azimuthal cumulants and Fourier harmonics as a function of the target saturation scale $Q_s$ and the transverse momentum. Additionally, we will perform a systematic study of the $N_c$ dependence of observables. In particular, we will point to key similarities and differences between the non-Abelian and Abelian versions of the model. We will also make predictions of yet to be measured symmetric cumulants for higher order Fourier harmonics.

Expectation values over the dipole correlators are computed in the McLerran-Venugopalan (MV) model~\cite{McLerran:1993ka,McLerran:1993ni}. This model includes coherent multiple scattering of the quarks in the projectile off the nuclear target. If we include at most two scatterings of the quarks, corresponding to the expansion of the Wilson lines to lowest nontrivial order, the expectation values correspond to the Glasma graph approximation. This approximation is applicable for $p_\perp > Q_s$. A model including quantum evolution of the Glasma graphs in the Color Glass Condensate (CGC) framework~\cite{Iancu:2003xm,Gelis:2010nm} was previously applied to successfully describe key features of azimuthal correlations for $p_\perp \geq Q_s$~\cite{Dumitru:2008wn,Dumitru:2010iy,Dusling:2012iga,Dusling:2012cg,Dusling:2012wy,Dusling:2013qoz,Dusling:2015rja}. We show that the Glasma graph correlators only produce positive values of the four-particle cumulant $c_2\{4\}$ and therefore do not correspond to a real $v_2\{4\}$. This result demonstrates that coherent multiple scattering, which is significant for $p_\perp \leq Q_s$, is an essential ingredient for the collectivity seen in our initial state framework.

The organization of the paper is as follow. We begin with the setup of our model in Sec.~\ref{sec:model_basics}.  In Sec.~\ref{sec:dipolecorr}, we discuss the algorithm for the computation of multi-dipole correlators. Results of our computations are presented in Sec.~\ref{sec:results}.  In  Sec.~\ref{sec:discussion}, we discuss the dependence of these results on the relative separations of the quarks in the projectile, on the number of color domains for varying $p_\perp$ windows and on the number of colors $N_c$. We contrast our results to those in the Glasma graph approximation. We briefly discuss the rapidity dependence on correlations in our model. In Sec.~\ref{sec:conclusion}, we conclude and discuss possible future directions of research. Details of the Glasma graph computations are presented in the \hyperref[app:gg]{Appendix}.

\section{Eikonal quark scattering from a nuclear target}
\label{sec:model}

\label{sec:model_basics}

We will discuss in this section a simple parton model description of proton-nucleus collisions. The incoming projectile consists of a collection of independent, nearly collinear, quarks that scatter off a dense nuclear target.  Spatial correlations within the classical field of the nucleus imprint themselves on the quarks as they scatter, resulting in nontrivial momentum space correlations between the originally uncorrelated quarks. These include correlations in their relative azimuthal angles.

We begin by considering the scattering of a fermion off a classical background field in the high energy limit~\cite{Bjorken:1970ah,Mueller:1994gb}.  The forward scattering amplitude for a fixed background field $A^-$ can be expressed as~\cite{Dumitru:2002qt}
\begin{eqnarray}
\left \langle q(\mathbf{q})_{out} | q(\mathbf{p})_{in} \right \rangle_{A^-} = \int d^2\bperp{x} \left[ U(x_\perp) -1 \right] e^{i(\mathbf{q}-\mathbf{p})\cdot \bperp{x}}\sim \mathcal{M}_{}(p,q)\,,
\label{eqn:amplitude}
\end{eqnarray}
where
\begin{eqnarray}
U(\bperp{x})&=&\mathcal{P}\text{exp}\Big(-ig \int dz^+{A^a}^-(\bperp{x},z^+)\,t^a \Big)\,
\label{eqn:wilsonline}
\end{eqnarray}
is the Wilson line in the fundamental representation at a transverse position $\bperp{x}$ and $\mathcal{P}$ denotes path ordering in the lightcone variable $x^+$.  The $-1$ in \Eq{eqn:amplitude} removes the ``no scattering" contribution wherein a quark passes through the target nucleus without having its color rotated by an Eikonal phase.  As the incoming partons all have transverse momentum of order $\lqcd$, and we are interested in $| \mathbf{p} | \gg \lqcd$, we will ignore this contribution in what follows.

The transverse spatial distribution of collinear quarks with transverse momenta $\mathbf{k}$ in the projectile is represented by the Wigner function $W_q(\mathbf{b},\mathbf{k})$~\cite{Lappi:2015vha,Lappi:2015vta}. The single inclusive distribution within this model can be expressed as
\begin{eqnarray}
\left\langle \frac{dN}{d^2\mathbf{p}} \right\rangle &=& \frac{1}{4\pi\Bp} \int d^2\mathbf{r} \int d^2\mathbf{b} \int \frac{d^2\mathbf{k}}{(2\pi)^2}  ~W_q(\mathbf{b},\mathbf{k}) e^{i (\mathbf{p}-\mathbf{k})\cdot \mathbf{r}}\left\langle D\left(\mathbf{b}+\frac{\mathbf{r}}{2},\mathbf{b}-\frac{\mathbf{r}}{2}\right) \right\rangle \,
\label{eqn:singleinclusive}
\end{eqnarray}
where the expectation value denotes an average over fields $A^-$ in the target, as for instance given by the MV model. For simplicity, we assume the Wigner function has the Gaussian form
\begin{eqnarray}
W_q(\mathbf{b},\mathbf{k})=\frac{1}{\pi^2} e^{-|\mathbf{b}|^2/\Bp}e^{-|\mathbf{k}|^2 \Bp}\,,
\label{eqn:wignergaussian}
\end{eqnarray}
where both the transverse momentum and spatial location of the quarks is determined by a single nonperturbative scale $\Bp$.
Unless otherwise mentioned, we will fix $\Bp=4~\text{GeV}^{-2}$~\cite{Kowalski:2006hc}, obtained from dipole model fits to HERA deep inelastic scattering data. We will discuss later how our results are affected by variations in the value of $\Bp$.

In \Eq{eqn:singleinclusive}, the function
\begin{eqnarray}
\label{eqn:dipoledef}
D(x,y)=\frac{1}{N_c}\text{Tr}\left[U(x)U^\dagger(y)\right]
\end{eqnarray}
denotes the dipole operator.   This operator encodes all orders in multiple gluon exchanges, as we will explicitly see in the calculations in Sec.~\ref{sec:dipolecorr}.
Performing the integration over the incoming quark momenta $\mathbf{k}$, \Eq{eqn:singleinclusive} can be simplified to read,
\begin{eqnarray}
\left\langle \frac{dN}{d^2\mathbf{p}} \right\rangle &=& \frac{1}{4\pi^3\Bp} \int d^2\mathbf{b}  \int d^2\mathbf{r} ~e^{-|\mathbf{b}|^2/\Bp}e^{-|\mathbf{r}|^2/4 \Bp}  ~e^{i\mathbf{p}\cdot\mathbf{r}} \langle D\left(\mathbf{b}+\frac{\mathbf{r}}{2},\mathbf{b}-\frac{\mathbf{r}}{2}\right) \rangle \,.
\label{eqn:singleinclusive}
\end{eqnarray}

The above framework can be extended to multiparticle production.  For $m$ incoming quarks in the projectile the $m$-particle inclusive spectrum can be expressed as
\begin{eqnarray}
\left\langle \frac{d^mN}{d^2\mathbf{p_1}\cdots d^2\mathbf{p_m}} \right\rangle \equiv\left\langle \frac{dN}{d^2\mathbf{p_1}}\cdots\frac{dN}{d^2\mathbf{p_m}}\right\rangle\,,
\label{eqn:twoparticle_inclusive}
\end{eqnarray}
where the expectation value denotes an average over classical configurations of the target in a single event and over all events. Since each of the single-particle distributions inside the average here is a gauge-dependent functional of the classical field, we caution the reader that these distributions are qualitatively different from the gauge invariant single-particle distributions employed in hydrodynamic computations. No such simple product of gauge invariant distributions can be written in our case; indeed, as discussed at length in the Appendix, the Feynman diagrams corresponding to \Eq{eqn:twoparticle_inclusive} are quantum interference diagrams.

\begin{figure}[ht]
\scalebox{.5}{\includegraphics[]{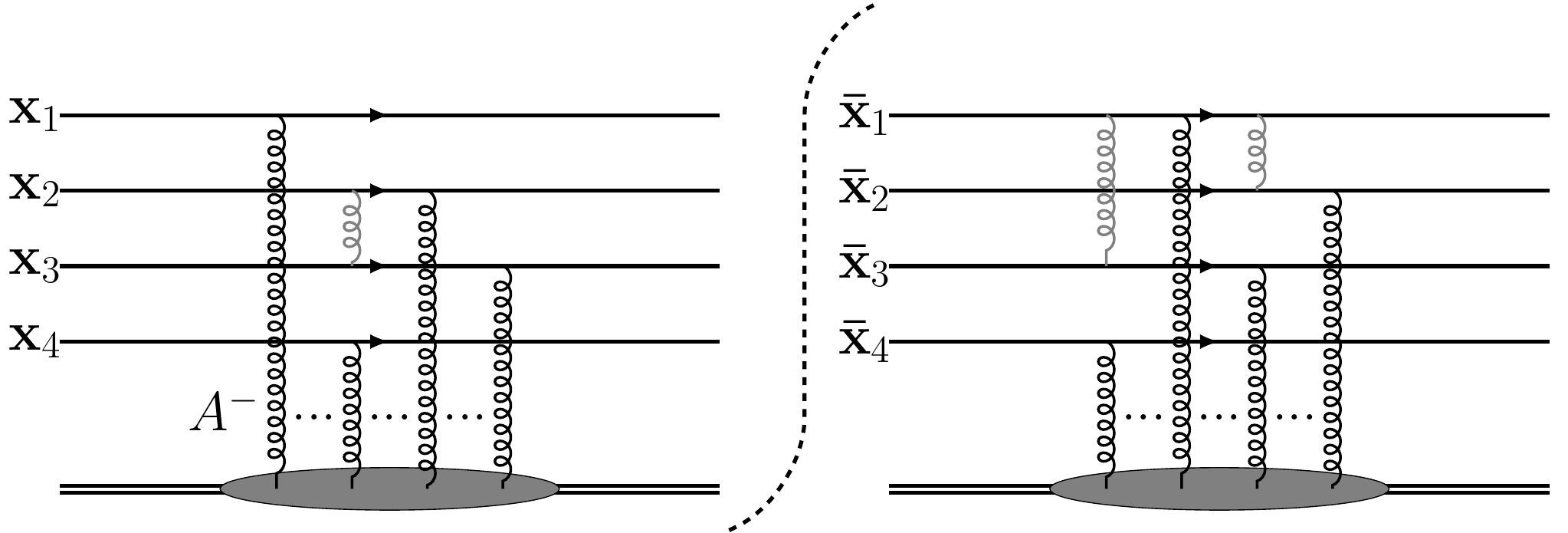}}
\caption{Diagrammatic representation of gluon exchanges between quarks in the amplitude (left) and complex-conjugate amplitude (right) and the target nucleus. The light-gray gluons show possible exchanges between quarks that would break the factorization used in \Eq{eqn:wignerfactorized}.  Correlations such as these might be generated via quantum evolution of the projectile and are not included in this work.  All allowed gluon exchanges between the quarks and the target are fully resummed.}
\label{fig:mult_scattering}
\end{figure}

On the other hand, the quarks comprising the projectile are uncorrelated,  with the $m$-quark Wigner function of the projectile factorizing into a product of single-quark Wigner functions,
\begin{eqnarray}
W(\mathbf{b_1},\mathbf{k_1},...,\mathbf{b_m},\mathbf{k_m})=W_{q}(\mathbf{b_1},\mathbf{k_1})\cdots\cdots W_q(\mathbf{b_m},\mathbf{k_m})\,.
\label{eqn:wignerfactorized}
\end{eqnarray}
Diagrammatically, this can be represented, as shown in \Fig{fig:mult_scattering}, as the multiple scattering between different quarks in the amplitude and complex conjugate amplitude and the target nucleus. In the strict dilute-dense limit in which we work, these Wigner functions are gauge invariant distributions of which the product form is assumed to survive color averaging. From \Eq{eqn:twoparticle_inclusive} and \Eq{eqn:wignerfactorized}, we arrive at the following compact form for the $m$-particle inclusive spectra:
\begin{eqnarray}
\label{eqn:multiplicity_kintegrated}
\left\langle\frac{d^{m} N}{d^2\mathbf{p_{1}}\cdots d^2\mathbf{p_{m}} } \right\rangle&=& \frac{1}{(4\pi^3 \Bp)^m} {\prod_{i=1}^m} \int d^2\mathbf{b_i}  \int d^2\mathbf{r_i}~e^{-|\mathbf{b_i}|^2/\Bp} e^{-|\mathbf{r_i}|^2/4\Bp}  e^{i\mathbf{p_{i}}\cdot \mathbf{r_i}}
\left< \prod_{j=1}^m D\left(\mathbf{b_j} + \frac{\mathbf{r_j}}{2},\mathbf{b_j} - \frac{\mathbf{r_j}}{2}\right) \right>\,.
\end{eqnarray}

Even though the above expression has a factorized form, it is highly nontrivial. Multiparticle correlations are generated via the expectation value over the classical fields of the target.  A primary focus of this work will be computing the correlation between four particles. In this case, the expectation value is over a product of four-dipole operators, each of which, as noted, is a trace of two lightlike Wilson lines.  The resulting expectation value is a function of eight transverse coordinates: four coordinates in the amplitude and four in the complex-conjugate amplitude.  This expectation value will be evaluated without approximation in the MV model.  We will see that large $N_c$ approximations and perturbative expansions (such as the Glasma graph approximation)
to such correlators are insufficient to capture the systematics of $pA$ data.

A shortcoming of our model is the oversimplified nature of the projectile. At high energies, gluon radiation dominates the small-$x$ component of the proton's wavefunction. These high parton densities become apparent when $Q_{s,T}/p_T \gtrsim 1$;
saturation model fits to HERA data conservatively suggest
that these effects become non-negligible around $x = 0.01$~\cite{Rezaeian:2012ji}.
However, depending on the transverse momentum range studied, the qualitative features we observe could persist to smaller values of $x$.
Furthermore, as the rapidity separation between quarks becomes larger than $\Delta y \gtrsim 1/\alpha_S$, quantum corrections will result in a decorrelation between partons.  A more quantitative discussion of the rapidity dependence is discussed in Sec.~\ref{sec:rap_dep}.

Quantum evolution will clearly break the factorized form of the Wigner function used in \Eq{eqn:wignerfactorized}. Furthermore, since gluons would dominate the scattering process, their interactions with the target would be represented as adjoint Wilson lines. Multiparticle production, in this dense-dense limit, has been addressed in previous work~\cite{Gelis:2008ad,Dusling:2009ni,Lappi:2009xa,Schenke:2015aqa,McLerran:2016snu,Kovner:2016jfp}. Multiparticle distributions can be obtained by solving the classical Yang-Mills equations in the presence of lightlike color sources corresponding to the projectile and the target. These source densities are each drawn from functional distributions of color charges, the evolution with energy of which is described by the JIMWLK equations~\cite{JalilianMarian:1997gr,Iancu:2000hn}. However, due to the numerical complexity of the simulations, this has been restricted thus far to two-particle correlations~\cite{Lappi:2009xa,Schenke:2015aqa}.  The success of our simple model in explaining many of the collective signatures seen in light--heavy-ion collisions should stimulate further development of classical Yang-Mills simulations.

One may note that \Eq{eqn:multiplicity_kintegrated} has the structure of an expectation value of a product of functions. If one interpreted these functions as ``single-particle'' distributions the form of \Eq{eqn:multiplicity_kintegrated} would be markedly similar to a hydrodynamic framework~\cite{Gronqvist:2016hym}. One may then conjecture that the results we show for $v_2\{4\}$ are simply a consequence of the functional form of~\Eq{eqn:multiplicity_kintegrated}. This turns out not to be the case. In our discussion of coherent multiple scattering versus Glasma graphs, we will
observe that, while both can be expressed in single-particle product form, one obtains negative four-particle azimuthal cumulants in the former case and positive valued cumulants in the latter\footnote{We thank Jean-Yves Ollitrault for a useful discussion on this point.}.

\section{Expectation values of multi-dipole correlators in the MV model}
\label{sec:dipolecorr}

In this section, we will compute the expectation value of four dipole operators in the MV model~\cite{McLerran:1993ka,McLerran:1993ni} of a nucleus at high energies.  Although we will only make use of dipole operators, the results also generate all allowed expectation values of eight Wilson lines at no additional computational cost.  The algorithm presented here to compute expectation values of lightlike Wilson line correlators can in principle be extended to higher-point functions.

In the MV model, classical gauge fields are described by solutions of the classical Yang-Mills equations,
\begin{eqnarray}
[D_\mu,F^{\mu\nu}]=\delta^{\nu-}\rho(\mathbf{x},x^+)\,,
\label{eqn:cym}
\end{eqnarray}
where $\rho$ denotes the classical color charge density in the nucleus. It is determined from the random Gaussian distribution satisfying
\begin{eqnarray}
\langle \rho_a(\bperp{x},x^+) \rho_b(\bperp{y},y^+) \rangle = \delta_{ab} \delta(x^+-y^+)\delta^{(2)}(\bperp{x}-\bperp{y})\,\mu^2(x^+)\,,
\label{eqn:rhorhocorrelator}
\end{eqnarray}
where  $\mu^2$ is the squared color charge density per unit area.  The above two-point function
can also be recast in terms of the gauge fields $A^{-}_a$ using
\begin{eqnarray}
A^-_a(\bperp{x},x^+)=g \int d^2\bperp{z} G(\bperp{x}-\bperp{z}) \rho_a(\bperp{z},x^+)\,,\qquad G(\mathbf{x}_\perp)=\int \frac{d^2\bperp{k}}{(2\pi)^2}\frac{e^{i\mathbf{k}_\perp \cdot \mathbf{x}_\perp}}{|\mathbf{k}_\perp |^2}\,,
\label{eqn:Apropagator}
\end{eqnarray}
where $G(\mathbf{x}_\perp)$ is the free gluon propagator in two dimensions~\cite{Gelis:2008rw}. One then obtains,
\begin{eqnarray}
g^2 \langle A^-_a(\bperp{x},x^+) A^-_b(\bperp{y},y^+) \rangle &=& g^4 \mu^2(x^+) \delta(x^+-y^+)\delta_{ab} \int d^2\bperp{z} G(\bperp{x}-\bperp{z}) G(\bperp{y}-\bperp{z})\nonumber\\
& \equiv&  \delta(x^+-y^+)\delta_{ab} L(\bperp{x},\bperp{y})\,.
\label{eqn:AAcorrelator}
\end{eqnarray}
The integral over the two-dimensional propagator is formally divergent and must be regulated at the nonperturbative scale $\lqcd$ with the result
\begin{eqnarray}
L(\mathbf{x}_{\perp},\mathbf{y}_{\perp})&=&(g^2\mu)^2 \int d^2 \bperp{z} G(\bperp{x} -\bperp{z} ) G(\bperp{x} -\bperp{y} ) \sim-\frac{(g^2\mu)^2 }{16\pi} | \bperp{x}-\bperp{y}|^2 ~\text{log}\left(\frac{1}{| \bperp{x}-\bperp{y}| \lqcd}  \right)\,.
\label{eqn:L_expanded}
\end{eqnarray}

Expectation values of multiple dipoles can be computed by expanding the path ordered exponential within the Wilson line to second order in the gauge field.  All possible pairwise contractions of the gauge fields are evaluated using \Eq{eqn:AAcorrelator}.  The result can be re-exponentied resulting in an expression valid to all orders in the gauge field~\cite{Kovner:2001vi,Fujii:2002vh}. There is also a formally equivalent graphical method~\cite{Blaizot:2004wv,Dominguez:2008aa}, which we will refer to when helpful.

When expanding out the path ordered exponential in the Wilson line, each gauge field represents a single gluon exchange with the target.
Each $\langle AA \rangle$ contraction is therefore equivalent to considering two gluon exchanges between two quarks (represented as Wilson lines) and the target nucleus.  Gluon exchanges can occur twice on the same quark, or on different quarks, and can also act on anti-quarks in the conjugate amplitude.

Following \cite{Blaizot:2004wv}, we will denote as $\mathcal{T}$ contributions arising from two gluon exchanges with the same quark and $\mathcal{N}$ as two-gluon exchange among different quarks. Since exchanges between the same Wilson line in $\mathcal{T}$ are color singlets, these contributions can be considered separately from the  $\mathcal{N}$ contributions. The final result for the expectation value of $m$-dipole operators can be expressed as a product of the two contributions,
\begin{eqnarray}
\left< D\cdots D \right> = \mathcal{T} ~\mathcal{N}\,.
\label{eqn:DDDD}
\end{eqnarray}

\subsection{Tadpole contribution}
\label{sec:tadpole}
\begin{figure}
\includegraphics[width=0.18\textwidth]{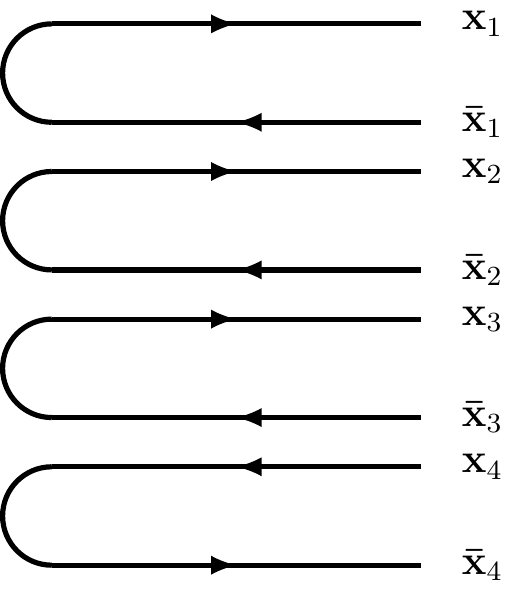}
\caption{Eight fundamental Wilson lines, pairwise connected. Coordinates denote transverse positions.}
\label{fig:fourdipoles_open}
\end{figure}

Gluon exchanges between the same Wilson line comprise the tadpole contribution $\mathcal{T}$. We take as our starting point eight Wilson lines having transverse positions $\mathbf{x}_1,\mathbf{\bar{x}}_1,\dots,\mathbf{x}_4,\mathbf{\bar{x}}_4$, where the $\mathbf{x}$ positions refer to quarks and the $\mathbf{\bar{x}}$ positions refer to anti-quarks. These dipoles must be connected in such a way to preserve the flow of color. Namely, quarks can only be connected to anti-quarks and vice versa. Without loss in generality, we connect at $x^+=-\infty$ quarks at $\mathbf{x}_i$ with anti-quarks at $\mathbf{\bar{x}}_i$, as shown in \Fig{fig:fourdipoles_open}.

\begin{figure}
\includegraphics[width=0.33\textwidth]{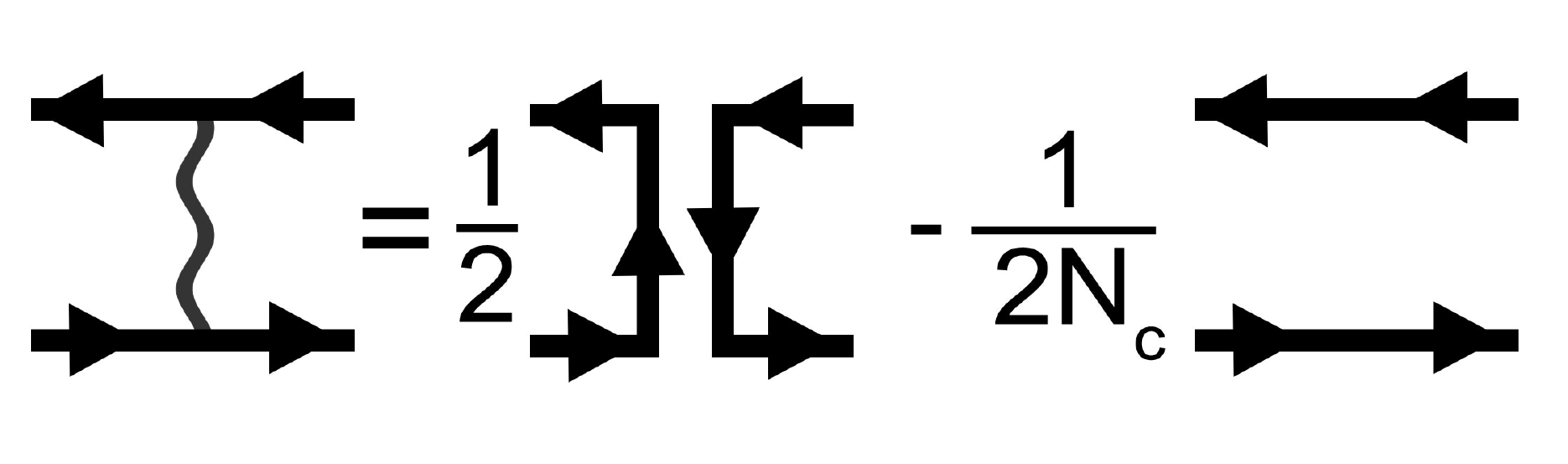}
\caption{The Fierz identity, \Eq{eqn:fierz}. }
\label{fig:fierz}
\end{figure}

\begin{figure}
\begin{subfigure}[t]{0.18\textwidth}
\includegraphics[width=\textwidth]{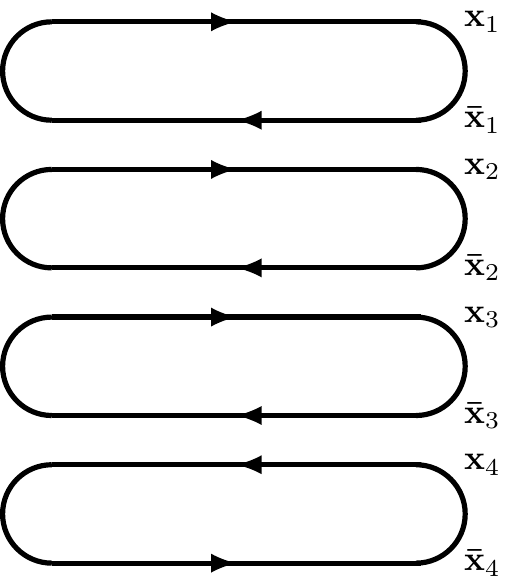}
\caption{Four dipoles.}
\label{fig:fourdipoles}
\end{subfigure}
\hspace{0.2cm}
\begin{subfigure}[t]{0.18\textwidth}
\includegraphics[width=\textwidth]{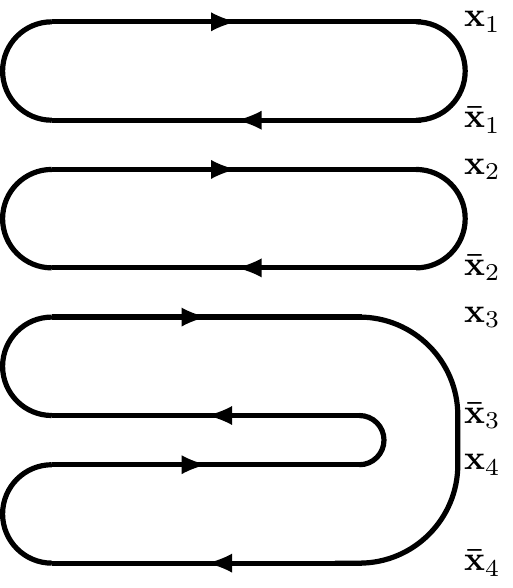}
\caption{Two dipoles and a quadrupole.}
\label{fig:dipoledipolequad}
\end{subfigure}
\hspace{0.2cm}
\begin{subfigure}[t]{0.18\textwidth}
\includegraphics[width=\textwidth]{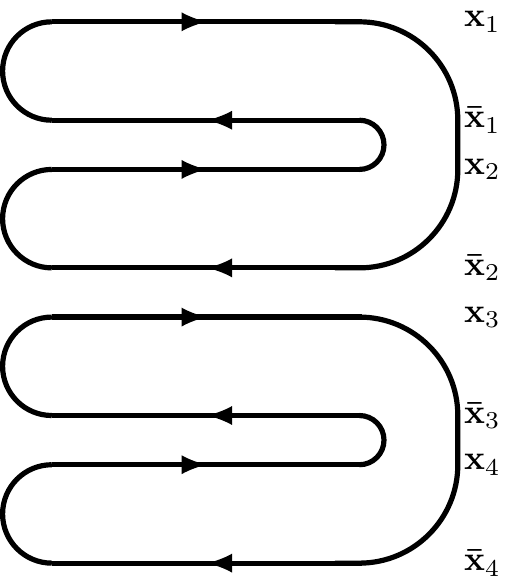}
\caption{Two quadrupoles.}
\label{fig:quadquad}
\end{subfigure}
\hspace{0.2cm}
\begin{subfigure}[t]{0.18\textwidth}
\includegraphics[width=\textwidth]{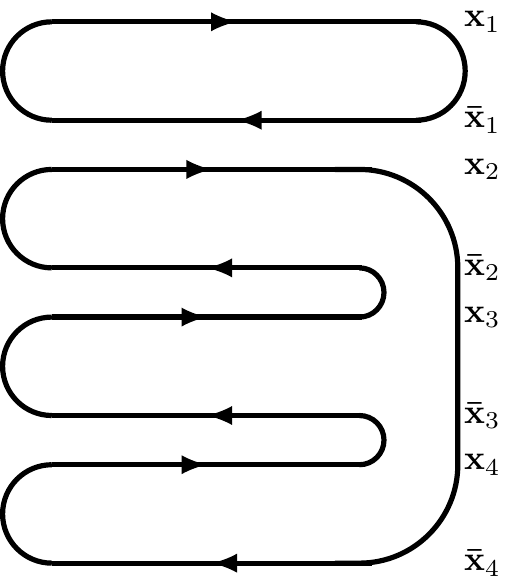}
\caption{One dipole and a sextupole.}
\label{fig:dipolesext}
\end{subfigure}
\hspace{0.2cm}
\begin{subfigure}[t]{0.18\textwidth}
\includegraphics[width=\textwidth]{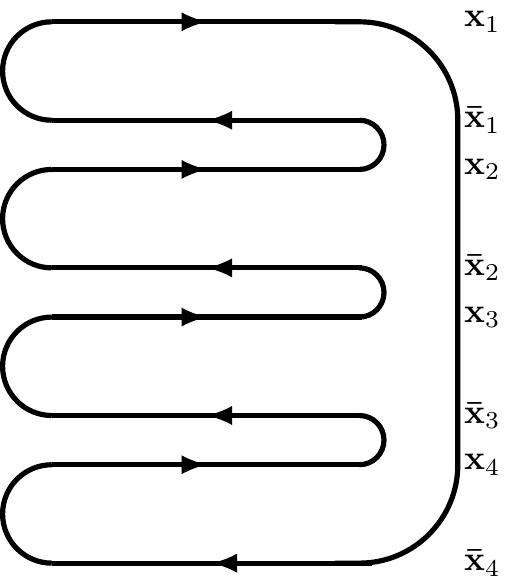}
\caption{An octupole.}
\label{fig:octupole}
\end{subfigure}
\caption{The five different topologies possible for eight Wilson lines. All possible permutations with respect to the given transverse coordinates are possible.}
\label{fig:eightwilsonlinetopo}
\end{figure}

This allows us to unambiguously define what we mean as a four-dipole configuration for the given positions\footnote{Physically however these positions are completely arbitrary.}, as shown in Fig.~\ref{fig:fourdipoles}. However, we can consider the $x^+=+\infty$ ends to be initially open ended.
To evaluate the tadpole contribution, we draw a gluon connecting at two arbitrary $x^+$ points on the same Wilson line of \Fig{fig:fourdipoles_open}. We then invoke the Fierz identity,
 \begin{eqnarray}
t^a_{ij}t^a_{kl}=\frac{1}{2}\delta_{il}\delta_{jk}-\frac{1}{2N_c}\delta_{ij}\delta_{kl},
\label{eqn:fierz}
\end{eqnarray}
which is given graphically by Fig.~\ref{fig:fierz}.

Performing this Fierz-ing, we see that the first term gives us a closed loop, which just is the trace of the identity matrix.  This gives a factor of $N_c$ and the dipole again.  The second term is just the original dipole. The result is then the original dipole multiplied with the Casimir factor $C_F=\frac{N_c^2-1}{2 N_c}$.

For the rest of the result, we must calculate pair-wise contractions of the gauge field $A^-_a$. As noted, for the MV model, these satisfy \Eq{eqn:AAcorrelator}. Every gluon exchange between the same (anti)-quark (same transverse position Wilson line) results in a factor of $-\frac{C_F}{2} L_{x_i x_i}$, where $L$ was defined in \Eq{eqn:L_expanded}, and the $1/2$ is on account of the fact that the two ends of the correlator are ordered in $x^+$ because they belong to the same Wilson line. The negative sign is from connecting a (anti)-quark with a (anti)-quark.

Summing $n$ such exchanges to the Wilson lines, the tadpole contribution can be expressed as
\begin{eqnarray}
\mathcal{T}=\text{exp}\Big({-\frac{C_F}{2} \displaystyle\sum_{i=1}^4\big(L_{x_i x_i}+L_{\bar{x}_i \bar{x}_i}\big)}\Big),
\label{eqn:tadpole}
\end{eqnarray}
where we now denote the transverse position arguments as subscripts for readability and differentiate between quarks in the amplitude ($x_i$) and anti-quarks in the complex-conjugate amplitude with an over-bar ($\bar{x}_i$).
This tadpole term is a color singlet and commutes with the terms we will derive next.

\subsection{Gluon exchange contribution}
\label{sec:gluon_exchange}
We shall now consider gluon exchanges between different Wilson lines. Our starting point is again the configuration shown in \Fig{fig:fourdipoles_open}. Closing the ends of the dipoles while preserving the color flow, we find that there are five distinct topologies. Using Fierz ordering as we did previously for the tadpole contribution, a gluon exchange between different Wilson lines in one topology can transform it into a different topology.

We now show how one obtains the five distinct topologies shown in Fig.~\ref{fig:eightwilsonlinetopo}. As a concrete example, we start with the four closed dipoles in Fig.~\ref{fig:fourdipoles}. Consider a gluon exchange between $\bar{x}_3$ and $x_4$. From the two terms that result from Fierz ordering, as shown in Fig.~\ref{fig:fierz}, the first term gives two dipoles and a quadrupole, as depicted in Fig.~\ref{fig:dipoledipolequad}, with a factor of $\frac{1}{2}$. A quadrupole, as depicted, is a distinct topological configuration corresponding to the trace over the product of Wilson lines at two distinct transverse spatial positions in the amplitude and at two such positions in the complex-conjugate amplitude.
The second term from Fierz-ing just returns the original four-dipole configuration shown in Fig.~\ref{fig:fourdipoles}, but now with the Fierz factor $\frac{1}{2N_c}$. Now taking this dipole-dipole-quadrupole configuration, consider further an exchange between $\bar{x}_1$ and $x_2$. This creates a quadrupole-quadrupole topology from the first term in the Fierz-ing, as depicted in Fig.~\ref{fig:quadquad}, and likewise the structure Fig.~\ref{fig:dipoledipolequad} from the second Fierz term, both terms with the appropriate Fierz prefactors.

If instead we had started with the dipole-dipole-quadrupole configuration and considered a gluon exchange between $\bar{x}_2$ and $x_3$, the first Fierz term would have resulted in a dipole-sextupole configuration, as depicted in Fig.~\ref{fig:dipolesext}. A sextupole, as depicted in Fig.~\ref{fig:dipolesext}, corresponds to a trace over the product of Wilson lines at three spatial positions in the amplitude and three in the complex-conjugate amplitude. An exchange between  $\bar{x}_1$ and $x_2$  in this dipole-sextupole topology results in an octupole (trace of eight Wilson lines--four in the amplitude and the other four in the complex-conjugate amplitude), depicted in Fig.~\ref{fig:octupole}, for the first Fierz term. The second term, as for the previous cases, gives back the original configuration, the dipole-sextupole one, with the appropriate Fierz prefactor. Thus, we see that multiple gluon exchanges continually generate, with each additional exchange, transitions between five topologically distinct configurations: (a) four-dipole, (b) dipole-dipole-quadrupole, (c) quadrupole-quadrupole, (d) dipole-sextupole, (e) octupole.

There is a transverse coordinate permutation degeneracy to these diagrams as well. The four-dipole is the only topology without this permutation degeneracy. For the dipole-dipole-quadrupole topology, there are $\binom{4}{2}=6$ possible permutations to close the $x^+=+\infty$ side of the eight pairwise connected Wilson lines we introduced previously. Similarly, there are $\frac{1}{2}\binom{4}{2}=3$ quadrupole-quadrupole permutations, $2\binom{4}{3}=8$ dipole-sextupole permutations, and $3!=6$ octopole permutations.
As a sanity check, since we have considered eight Wilson lines connected pairwise at one end ($x^+=-\infty$), the sum of these permutations degeneracies agrees with the total $4!=24$ possible different contractions of the $x^+=+\infty$ side, which is dictated from the fact that quark (anti-quark) Wilson lines can only connect with anti-quark (quark) Wilson lines.

It should be clear from our discussion that we have a closed system of 24 configurations whereby each of these are transformed, through gluon exchanges and Fierz-ing, into other configurations in this system. We can express the 24 possible configurations as elements of a basis characterizing the four-dipole system. Starting with this initial condition, where all other configurations are set to zero, one can construct a 24 by 24 transformation matrix $\mathbf{M}$ that transforms one set of basis elements to another with each gluon exchange and subsequent Fierz-ing. We can then deduce, either analytically~\cite{Kovner:2001vi,Fujii:2002vh,Fukushima:2007dy} or diagramatically~\cite{Blaizot:2004wv,Dominguez:2008aa}, what factors (via Fierz) are picked up in going from a basis element $\alpha$ to basis element $\beta$, which then define the elements $M_{\alpha \beta}$ of the matrix.

To understand how one fills in the arrays of this matrix, consider a  path ordered exponential for the Wilson line
\begin{eqnarray}
U(\bperp{x})&=&\mathcal{P}\text{exp}\Big(ig \int^\xi dz^+A^-(z^+,\bperp{x}) \Big) \simeq V(\bperp{x})(1+igA^-_a(\xi,\bperp{x})t^a+...) \,,
\label{eqn:wilsonline_expanded}
\end{eqnarray}
where we expanded out the last infinitesimal slice in rapidity, and $V(\bperp{x})$ is a redefinition of the original Wilson line excluding this last infinitesimal slice. Substituting this last expression in the dipole operator, we obtain
\begin{eqnarray}
\langle D(\bperp{x},\bperp{\bar{x}})_U\rangle &=& \frac{1}{N_c}\langle \text{tr}(U(\bperp{x})U^\dagger(\bperp{\bar{x}})\rangle = \langle D(\bperp{x},\bperp{\bar{x}})_V\rangle + g^2 \langle A^-_a(\bperp{x})A^-_b(\bperp{\bar{x}})\rangle\frac{1}{N_c}\langle\text{tr}\left(V(\bperp{x})t^a t^b V^\dagger(\bperp{\bar{x}})\right)\rangle\,.
\label{eqn:dipole_expanded}
\end{eqnarray}
Here we have made use of the locality in rapidity of correlators in the MV model.

Using \Eq{eqn:AAcorrelator}, we can express $\langle D_U \rangle$ in the l.h.s in terms of $\langle D_V \rangle$ alone on the r.h.s.. Iterating this expression for each slice in rapidity, one obtains the exponentiated expression
\begin{eqnarray}
\langle D(\bperp{x},\bperp{\bar{x}})_U\rangle &=& e^{C_f L(\bperp{x},\bperp{\bar{x}})}\langle D(\bperp{x},\bperp{\bar{x}})_V\rangle\,.
\label{eqn:dipole_expanded_sum}
\end{eqnarray}
While this is a simple example, an identical procedure can be followed for multiple dipoles, or any higher order configuration, resulting from an initial four-dipole configuration.
Another example is the well-known two-dipole result~\cite{Blaizot:2004wv,Dominguez:2008aa,Kovner:2001vi,Fujii:2002vh}.  In this case, following the procedure outlined above, it is straightforward to show that
\begin{eqnarray}
\langle D_{x_1\bar{x}_1}D_{x_2 \bar{x}_2}\rangle_U&\simeq&\alpha_{x_1 \bar{x}_1 x_2 \bar{x}_2}\langle D_{x_1\bar{x}_1}D_{x_2 \bar{x}_2}\rangle_V\nonumber \\
&+&\beta_{x_1 \bar{x}_2 x_2 \bar{x}_1}\langle Q_{x_1 \bar{x}_2 x_2 \bar{x}_1}\rangle_V
\end{eqnarray}
where $Q= {\rm Tr}\left(V(x_1)V^\dagger(\bar{x}_2)V(x_2)V^\dagger(\bar{x}_1)\right)/N_c$ is the quadrupole configuration. Since the only two configurations for two dipoles are the dipole-dipole and quadrupole, we can then write this as a matrix equation:
\begin{eqnarray}
\begin{pmatrix}
\langle D_{x_1\bar{x}_1}D_{x_2 \bar{x}_2}\rangle \\
\langle Q_{x_1 \bar{x}_2 x_2 \bar{x}_1} \rangle
\end{pmatrix}_U
=
\begin{pmatrix}
\alpha_{x_1 \bar{x}_1 x_2 \bar{x}_2} & \beta_{x_1 \bar{x}_2 x_2 \bar{x}_1}\\
\beta_{x_1 \bar{x}_1 x_2 \bar{x}_2} & \alpha_{x_1 \bar{x}_2 x_2 \bar{x}_1}
\end{pmatrix}_U
\begin{pmatrix}
\langle D_{x_1 \bar{x}_1}D_{x_2 \bar{x}_2}\rangle \\
\langle Q_{x_1 \bar{x}_2 x_2 \bar{x}_1} \rangle
\end{pmatrix}_V
\end{eqnarray}
where $\alpha$ and $\beta$ are simple functions of $L(\bperp{x},\bperp{y})$ given in Ref.~\cite{Blaizot:2004wv}.

For a further example, if we consider four dipoles, then it is only possible via one gluon exchange to get to the six possible dipole-dipole-quadrupole configurations or stay in the four-dipole configuration. One obtains
\begin{eqnarray}
\langle D_{x_1 \bar{x}_1}D_{x_2 \bar{x}_2}D_{x_3 \bar{x}_3}D_{x_4 \bar{x}_4}\rangle_U&\simeq&\alpha_{x_1 \bar{x}_1 x_2 \bar{x}_2 x_3 \bar{x}_3 x_4 \bar{x}_4}\langle D_{x_1\bar{x}_1}D_{x_2 \bar{x}_2}D_{x_3 \bar{x}_3}D_{x_4 \bar{x}_4}\rangle_V+\beta_{x_1 \bar{x}_2 x_2 \bar{x}_1}\langle Q_{x_1 \bar{x}_2 x_2 \bar{x}_1} D_{x_3 \bar{x}_3}D_{x_4 \bar{x}_4}\rangle_V \nonumber \\ &+& \beta_{x_1 \bar{x}_2 x_2 \bar{x}_1}\langle Q_{x_1 \bar{x}_3 x_3 \bar{x}_1} D_{x_2 \bar{x}_2}D_{x_4 \bar{x}_4}\rangle_V+\beta_{x_1 \bar{x}_4 x_4 \bar{x}_1}\langle Q_{x_1 \bar{x}_4 x_4 \bar{x}_1} D_{x_2 \bar{x}_2}D_{x_3 \bar{x}_3}\rangle_V \nonumber \\ &+& \beta_{x_2 \bar{x}_3 x_3 \bar{x}_2}\langle Q_{x_2 \bar{x}_3 x_3 \bar{x}_2} D_{x_1 \bar{x}_1}D_{x_4 \bar{x}_4}\rangle_V + \beta_{x_2 \bar{x}_4 x_4 \bar{x}_2} \langle Q_{x_2 \bar{x}_4 x_4 \bar{x}_2} D_{x_1 \bar{x}_1}D_{x_3 \bar{x}_3}\rangle_V \nonumber \\ &+& \beta_{x_3 \bar{x}_4 x_4 \bar{x}_3} \langle Q_{x_3 \bar{x}_4 x_4 \bar{x}_3} D_{x_1 \bar{x}_1}D_{x_2 \bar{x}_2}\rangle_V \, .
\end{eqnarray}
We can repeat this for all topologies to obtain a similar matrix as for the two-dipole case.

\begin{figure*}
\includegraphics[width=0.75\textwidth]{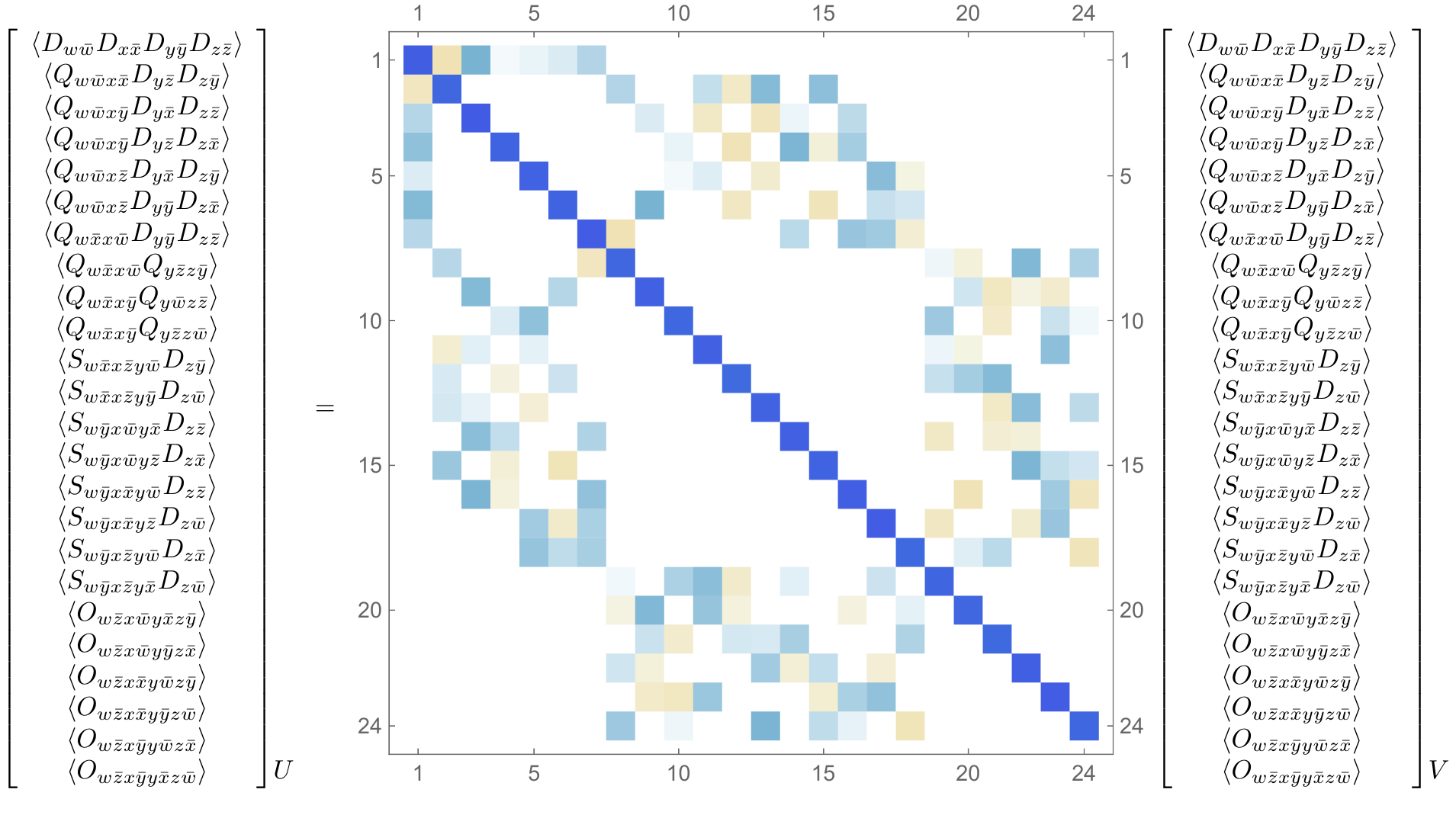}
\caption{The basis of configurations resulting for eight Wilson lines, pairwise connected, resulting in dipoles, quadrupole, sextupoles, and octupoles. Coordinates denote transverse positions.}
\label{fig:fourdipole_matrix}
\end{figure*}

For a small number of dipoles, this procedure can be carried out efficiently by hand, and the eigenvalues of the matrix can be even computed analytically. However for larger numbers of dipoles, this becomes cumbersome. For example, for the four dipoles we have been considering, we must compute $7!$ gluon exchanges for each of the 24 basis elements, totaling 120,960 computations. Fortunately, since the algorithm suggested by our exercises is quite straightforward, it is very tractable to determine the elements of the 24 by 24 matrix and compute their eigenvalues on a computer. This work made extensive use of the GNU Scientific Library~\cite{Gough:2009:GSL:1538674} and the EXPOKIT software package \cite{EXPOKIT}. The matrix is illustrated schematically by Fig.~\ref{fig:fourdipole_matrix}; as it indicates, the elements of the matrix are relatively sparse but must nevertheless be diagonalized numerically.  Our algorithm can be generalized to a larger number of dipoles of more complex topologies. It is therefore potentially useful for computations of many-body final states in high energy QCD where $n$-tupoles of lightlike Wilson lines are ubiquitous.

As a final remark on this matrix computation, we note that in the large $N_c$ limit this problem becomes more tractable~\cite{Dominguez:2012ad,Shi:2017gcq}. In Ref.~\cite{Dominguez:2012ad}, it was shown that for large $N_c$ only dipoles and quadrupoles contribute to high energy QCD
processes. However the azimuthal cumulants themselves vanish at large $N_c$, so in order to compute these quantities a finite $N_c$ calculation is necessary.

\subsection{Result for product of four-dipole correlators}
Now that we have explained how to calculate the matrix for one gluon exchange at a single slice in rapidity, we can compute how the basis vector space of configurations evolves after an infinite number of gluon exchanges. More generally, after $n$ gluon exchanges, we have the vector
\begin{eqnarray}
\vec{\mathcal{N}}_n=\text{a}_n \vec{\mathcal{N}}^{(\text{a})}+\text{b}_n \vec{\mathcal{N}}^{(\text{b})}+...+\text{x}_n \vec{\mathcal{N}}^{(\text{x})}\,,
\end{eqnarray}
where $\vec{\mathcal{N}}^{(a)}$ through $\vec{\mathcal{N}}^{(x)}$ refer to individual basis vectors in the 24-dimensional basis space. This then evolves according to
\begin{eqnarray}
\vec{\mathcal{N}}_{n+1}= \mathbf{M} \,\vec{\mathcal{N}}_{n}\,.
\end{eqnarray}
One can rewrite this expression as the matrix equation
\begin{eqnarray}
\begin{pmatrix}
\text{a}_{n+1} \\
\text{b}_{n+1} \\
... \\
\text{x}_{n+1}
\end{pmatrix}
=\mathbf{M}
\begin{pmatrix}
\text{a}_{n} \\
\text{b}_{n} \\
... \\
\text{x}_{n}
\end{pmatrix}.
\label{eqn:Mdef}
\end{eqnarray}
The setup of the computation thus far is general.  However, as previously stated, we wish to begin with an initial condition that is the closed four-dipole configuration.  This is done by setting the initial condition $\vec{\mathcal{N}}_0=\vec{\mathcal{N}}^{(a)}$ (the four-dipole configuration).  Since we know how to compute additional gluon exchanges from \Eq{eqn:Mdef}, we need to multiply our initial condition $\vec{\mathcal{N}}_0=\vec{\mathcal{N}}^{(a)}$ by $\mathbf{M}$ $n$ times. This is, however, just a compact way to write all possible configurations with appropriate factors after $n$ gluon exchanges from the starting point of four closed dipoles.  The result, after all orders in gluon exchanges, is simply
\begin{eqnarray}
\vec{\mathcal{N}}=\sum_{n=0}^\infty \frac{1}{n!} \vec{\mathcal{N}}_n=\sum_{n=0}^\infty \frac{1}{n!} \mathbf{M}^n\vec{\mathcal{N}}_0= e^{\mathbf{M}}\vec{\mathcal{N}}_0\,.
\end{eqnarray}
Lastly, to compute the expectation value over the product of four-dipole operators, we need to sum over each of the elements of the 24-dimensional column vector with the respective $N_c$ weights of each of the $n$-tupoles. The four-dipole configuration in this formulation has weight unity. The dipole-dipole-quadrupole configuration comes with $1/N_c$, both quadrupole-quadrupole and dipole-sextupole configurations with $1/N_c^2$, and octupoles with $1/N_c^3$. The sum over the 24-dimensional basis can then be written as the scalar product of the corresponding row vector and the column vector $\vec{\mathcal{N}}$:
\begin{eqnarray}
\mathcal{N}=(1, 1/N_c,.., 1/N_c, 1/N_c^2,..., 1/N_c^2,1/N_c^3, ..., 1/N_c^3) \,\vec{\mathcal{N}}\,.
\end{eqnarray}
Here the $\cdots$ denote the different permutations of each of the five configurations in Fig.~\ref{fig:eightwilsonlinetopo}. With this expression for $\mathcal{N}$, the combined gluon exchange and tadpole contributions in \Eq{eqn:tadpole} can be written as
\begin{eqnarray}
\langle DDDD \rangle &=& \mathcal{T} ~\mathcal{N}\,.
\label{eqn:FourDipoleResult}
\end{eqnarray}
As we noted, analytical expressions for these quantities are too cumbersome to compute. However, with the procedure outlined, they can be computed numerically and utilized to compute the four-particle correlation functions we shall discuss further in the next section.

\subsection{Abelian limit}
\label{app:abelian_limit}

We will consider here the computation of the $m$-dipole expectation value in an Abelianized version of the MV model. In the $U(1)$ theory, the Wilson line again represents the multiple scattering of a charged particle off a classical field~\cite{Bjorken:1970ah}:
\begin{equation}
U(\bperp{x})=\mathcal{P}\exp\left(-ie \int dz^+{A}^-(\bperp{x},z^+) \right)\,.
\label{eqn:mvwilsonlineqed}
\end{equation}
However, here the Wilson line is a scalar valued function, not an $SU(N_c)$ valued matrix as in the non-Abelian case; this simplifies computations enormously. Expectation values of multiple Wilson lines can be evaluated analogously to the non-Abelian case by first expanding each exponential to second order in the gauge field, followed by replacing all pairwise contractions of the gauge field with the Gaussian expectation value, as in~\Eq{eqn:AAcorrelator}:
\begin{eqnarray}
e^2 \langle A^-(\bperp{x},x^+) A^-(\bperp{y},y^+) \rangle &=& e^4 \mu^2 \delta(x^+-y^+) \int d^2\bperp{z} G(\bperp{x}-\bperp{z}) G(\bperp{y}-\bperp{z}) \equiv \delta(x^+-y^+) L(\bperp{x},\bperp{y})\,.
\label{eqn:AAcorrelator_abelian}
\end{eqnarray}

The dipole expectation value is then
\begin{eqnarray}
\langle D(\bperp{x},\bperp{\bar{x}})\rangle=\langle U(\bperp{x})U^*(\bperp{\bar{x}})\rangle=\exp\left[L(\bperp{x},\bperp{\bar{x}})\right]\,,
\end{eqnarray}
and the correlator of two dipoles can be expressed as
\begin{eqnarray}
\langle D(\bperp{x},\bperp{\bar{x}})D(\bperp{y},\bperp{\bar{y}})\rangle= \langle U(\bperp{x})U^*(\bperp{\bar{x}})U(\bperp{y})U^*(\bperp{\bar{y}})\rangle&=&\text{exp}\left( L_{{x},{\bar{x}}}+L_{{y},{\bar{y}}}+L_{{x},{\bar{y}}}+L_{{y},{\bar{x}}}-L_{{x},{y}}-L_{{\bar{x}},{\bar{y}}} \right)\,.
\end{eqnarray}
Similarly, for four dipoles, one obtains
\begin{equation}
\begin{split}
\langle D(\bperp{x},\bperp{\bar{x}})&D(\bperp{y},\bperp{\bar{y}})D(\bperp{z},\bperp{\bar{z}})D(\bperp{w},\bperp{\bar{w}})\rangle  \\
&=\text{exp} \big(L_{{x},{\bar{x}}} + L_{{y},{\bar{y}}} + L_{{z},{\bar{z}}} + L_{{w},{\bar{w}}} + L_{{x},{\bar{y}}} + L_{{x},{\bar{z}}} + L_{{x},{\bar{w}}} +L_{{y},{\bar{x}}} + L_{{y},{\bar{z}}} + L_{{y},{\bar{w}}} + L_{{z},{\bar{x}}} + L_{{z},{\bar{y}}}   \\
&\phantom{=} \qquad \, \, + L_{{z},{\bar{w}}}  + L_{{w},{\bar{x}}}  + L_{{w},{\bar{y}}}   + L_{{w},{\bar{z}}} - L_{{x},{y}} - L_{{x},{z}} - L_{{x},{w}} - L_{{y},{z}} - L_{{y},{w}} - L_{{z},{w}}  - L_{{\bar{x}},{\bar{y}}}   \\
&\phantom{=} \qquad \, \, - L_{{\bar{x}},{\bar{z}}} - L_{{\bar{x}},{\bar{w}}} - L_{{\bar{y}},{\bar{z}}} - L_{{\bar{y}},{\bar{w}}}  - L_{{\bar{z}},{\bar{w}}} \big)\,.
\end{split}
\end{equation}
Higher-point correlators can be found similarly by summing over all combinations of two-point functions; a negative sign is introduced for combinations of two quarks or two anti-quarks.

\section{Results}
\label{sec:results}

We will now discuss the calculation of $m$-particle production using the expression in \Eq{eqn:multiplicity_kintegrated}.  The inputs into this expression include the parameter $\Bp$ representing the transverse area of the projectile and the function $L(\mathbf{x}_{\perp},\mathbf{y}_{\perp})$ that represents the correlation of gauge field configurations arising from a Gaussian distribution of color sources.
In the MV model, the quantity $L(\mathbf{x}_{\perp},\mathbf{y}_{\perp})$ is related to the expectation value of the dipole operator through the expression
\begin{eqnarray}
\label{eqn:dipoleop}
\langle D(\bperp{x},\bperp{y})\rangle=e^{C_F L(\bperp{x},\bperp{y})} \, .
\end{eqnarray}
In this work, we will use the functional form
 \begin{eqnarray}
L(\mathbf{x}_{\perp},\mathbf{y}_{\perp})&=&-\frac{(g^2\mu)^2 }{16\pi} | \bperp{x}-\bperp{y}|^2 ~\text{log}\left(\frac{1}{| \bperp{x}-\bperp{y}| \Lambda} +e \right) \, .
\label{eqn:L_expanded2}
\end{eqnarray}
For the infrared cutoff in the model, we will use the value $\Lambda=0.241~\text{GeV}$. This value is obtained from parametrizations of the dipole amplitude used in dipole model computations of structure functions that are fit to the HERA deep inelastic scattering  $ep$ data~\cite{Albacete:2010sy,Lappi:2013zma}. We have checked that the results are qualitatively unchanged within a physically reasonable range of values of $\Lambda$.

Following Ref. \cite{Lappi:2013zma}, a model independent saturation scale is defined through the relation
\begin{equation}
\left<D\left(| \bperp{x}-\bperp{y}|^2=2/\Qs^2\right)\right>=e^{-1/2}\,.
\end{equation}
For the remainder of this work, we will specify values of $\Qs^2$ rather than $(g^2\mu)^2$.  We should point out that the mapping between
$(g^2\mu)^2$ and $\Qs^2$ contains an explicit dependence on $C_F$, as is transparent from \Eq{eqn:dipoleop}.  When we compare results at various $N_c$, this scaling with $C_F$ is taken into account. The exception is the $U(1)$ case, where we take $C_F=4/3$ when relating $Q_s^2$ to $(g^2\mu)^2$.

We stress that the saturation scale $\Qs$ here is that of the target nucleus.  There is no analogous saturation scale of the projectile in the model we are considering.  This is a consequence of the simplicity of our treatment of the projectile's constituents, which are comprised of nearly collinear uncorrelated quarks alone.  The corresponding average multiplicity per interaction is unity. This can be seen by explicit integration of \Eq{eqn:multiplicity_kintegrated} and can be contrasted with the expression for the multiplicity found in $k_\perp$-factorization~\cite{Dumitru:2001ux}
\begin{eqnarray}
N_{\rm mult}\sim Q_{s,p}^2 S_\perp \log\left(\frac{Q_{s,T}^2}{Q_{s,p}^2}\right) \, ,
\end{eqnarray}
where $Q_{s,p}$ and $Q_{s,T}$ are the saturation scales of the projectile and target respectively.  The transverse overlap area $S_\perp$ can be identified with the projectile area $\Bp$ in our model.  The absence of a projectile saturation momentum is the biggest shortcoming of the above framework.  The equivalent scale controlling the momenta of the incoming quarks is $1/\Bp$, which is held fixed. While the fact that our model and the experimental data seem to both be independent of the multiplicity (at least qualitatively) may be suggestive of a common physical origin, it is far from clear that this will hold for a more realistic model of the projectile.

In what follows, we will primarily plot quantities as a function of the target saturation scale $\Qs$.  The target saturation scale is a function of both Bjorken $x$ and the impact parameter.  As discussed above, the multiplicity is a logarithmic function of $\Qs$.  Instead, $\Qs$ is better thought of as a proxy for the energy of the collision. In the CGC picture of high energy QCD, $Q_s$ grows with the center-of-mass energy.

As a final remark, we stress that we only expect to make a qualitative comparison with data.  In addition to the shortcomings of the model addressed above, our final state distributions are for quarks and not for hadrons. Any correlations computed in this model will be reduced through a variety of effects, such as fragmentation, and quantum evolution of parton distributions. They therefore provide an upper limit for azimuthal correlations in initial state frameworks.

\subsection{Multiparticle azimuthal cumulants and harmonics}
\label{sec:exp_cumulant}

Multiparticle correlations carry a wealth of information on the dynamics of the colliding system. For reviews, see for instance, Refs.~\cite{Dusling:2015gta,Schlichting:2016sqo}.  Azimuthal correlations, in particular, are sensitive measures of collective dynamics in heavy-ion collisions. For systems undergoing collective flow, one can define the $n$th-Fourier harmonic coefficients
\begin{equation}
v_n=\left< e^{in\left(\phi-\Psi_R\right)}\right> \,,
\end{equation}
where $\phi$ is the azimuthal angle of a produced particle and $\Psi_R$
is the angle of a reaction plane determined by the collective geometry of particles produced in the collision. The determination of a
suitable reaction plane may be sensitive to
a variety of so-called {\em non-flow} contributions such as resonance decays, to give one example.  The effect of these {\em non-flow}
correlations can be minimized using cumulant method~\cite{Borghini:2001vi}, which is now widely used in experimental studies of multiparticle correlations. Two- and four-particle cumulants are defined as
\begin{eqnarray}
c_n\{2\}&=&\langle  e^{in(\phi_1-\phi_2)} \rangle \label{eqn:cn2} \\
c_n\{4\}&=&\langle  e^{in(\phi_1+\phi_2-\phi_3-\phi_4)} \rangle - 2 \langle  e^{in(\phi_1-\phi_2)}\rangle^2 \,,\label{eqn:cn4}
\label{eqn:intrinsic-cumulants}
\end{eqnarray}
where the average $\langle \cdots \rangle$ in $c_n\{m\}$ is the average of all possible combinations of $m$ particles in an event, followed by an averaging over all events \cite{Borghini:2001vi,Dumitru:2014yza,Khachatryan:2015waa}.

The cumulants above can be expressed in terms of the $m$-particle inclusive distribution by first defining a quantity $\kappa_n\{m\}$ as
\begin{eqnarray}
\kappa_n\{m\} \equiv \int  \prod_{i=1}^m  d^2\mathbf{p_{i}}\, e^{i n(\phi_1+\cdots+\phi_{m/2-1}-\phi_{m/2}-\cdots -\phi_{m})} \left< \frac{d^m N}{\prod_{i=1}^m d^2\mathbf{p_{i}}} \right> \,,
\label{eqn:kappa}
\end{eqnarray}
corresponding to the $n^{th}$ harmonic of the $m$-particle distribution.  The averages can be written as a ratio
of the $n^{th}$ azimuthal angle moment of the $m$-particle inclusive distribution normalized by the zeroth moment (the $m$-particle inclusive multiplicity):
\begin{eqnarray}
\left< e^{i n(\phi_1+\cdots+\phi_{m/2-1}-\phi_{m/2}-\cdots\phi_{m})} \right> = \frac{\kappa_{n}\{m\}}{\kappa_{0}\{m\}} \,.
\label{eqn:azimuth_corr}
\end{eqnarray}
Fourier coefficients are then defined from the above cumulants as
\begin{eqnarray}
v_n\{2\}&=&(c_n\{2\})^{1/2}\,, \\
v_n\{4\}&=&(-c_n\{4\})^{1/4}\,.
\label{eqn:v24ref}
\end{eqnarray}

The motivation for the above definitions becomes transparent under the assumption that the $m$-particle distribution factorizes into a product of single-inclusive distributions correlated with each other only through the event plane angle. This is indeed what one would expect if the system undergoes hydrodynamic flow. In such a framework, the two-particle Fourier harmonic $v_n\{2\}$ would, as above, be the square root of the corresponding two-particle azimuthal angle cumulant, the four-particle Fourier harmonic $v_n\{4\}$, the fourth root of the four particle azimuthal angle cumulant, and so on. The negative sign in the latter case is appropriate because one anticipates intrinsic four-particle angular correlations to be subdominant relative to the square of the two-particle cumulant in \Eq{eqn:intrinsic-cumulants}.  While the observation of $m$-particle Fourier harmonics is suggestive of some form of collective behavior, we will argue instead that it can result from any physical mechanism where higher cumulants are suppressed relative to the mean and variance of the distribution.

\begin{figure}
\begin{minipage}{.45\textwidth}
\includegraphics[width=1.0\textwidth]{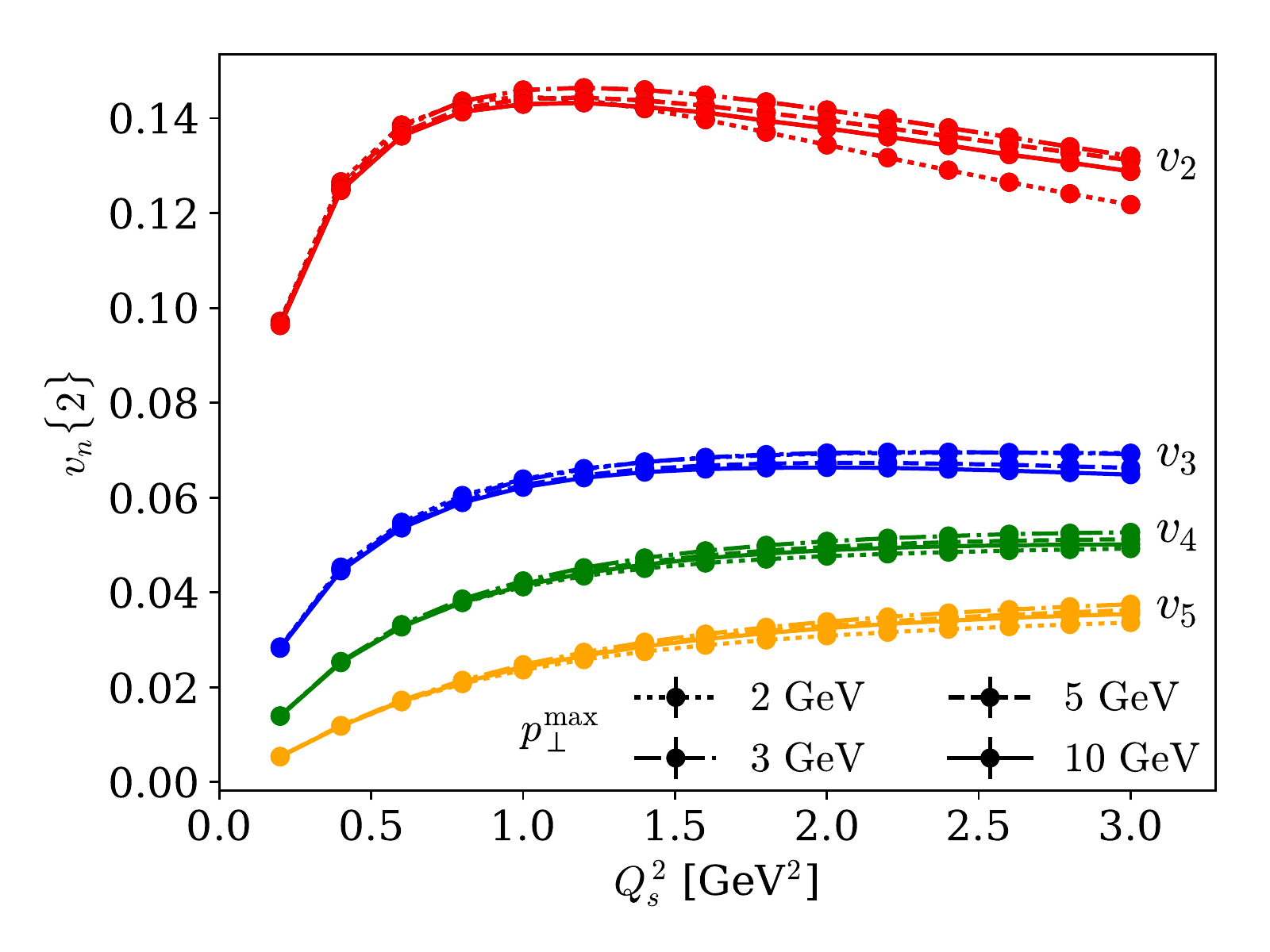}
\caption{Integrated $v_2\{2\}$ as a function of the maximum integrated momenta, $\pmax$, for various Fourier harmonic $n$, as a function of $Q_s^2$.}
\label{fig:vn2_ptmax}
\end{minipage}
\begin{minipage}{.45\textwidth}
\includegraphics[width=1.0\textwidth]{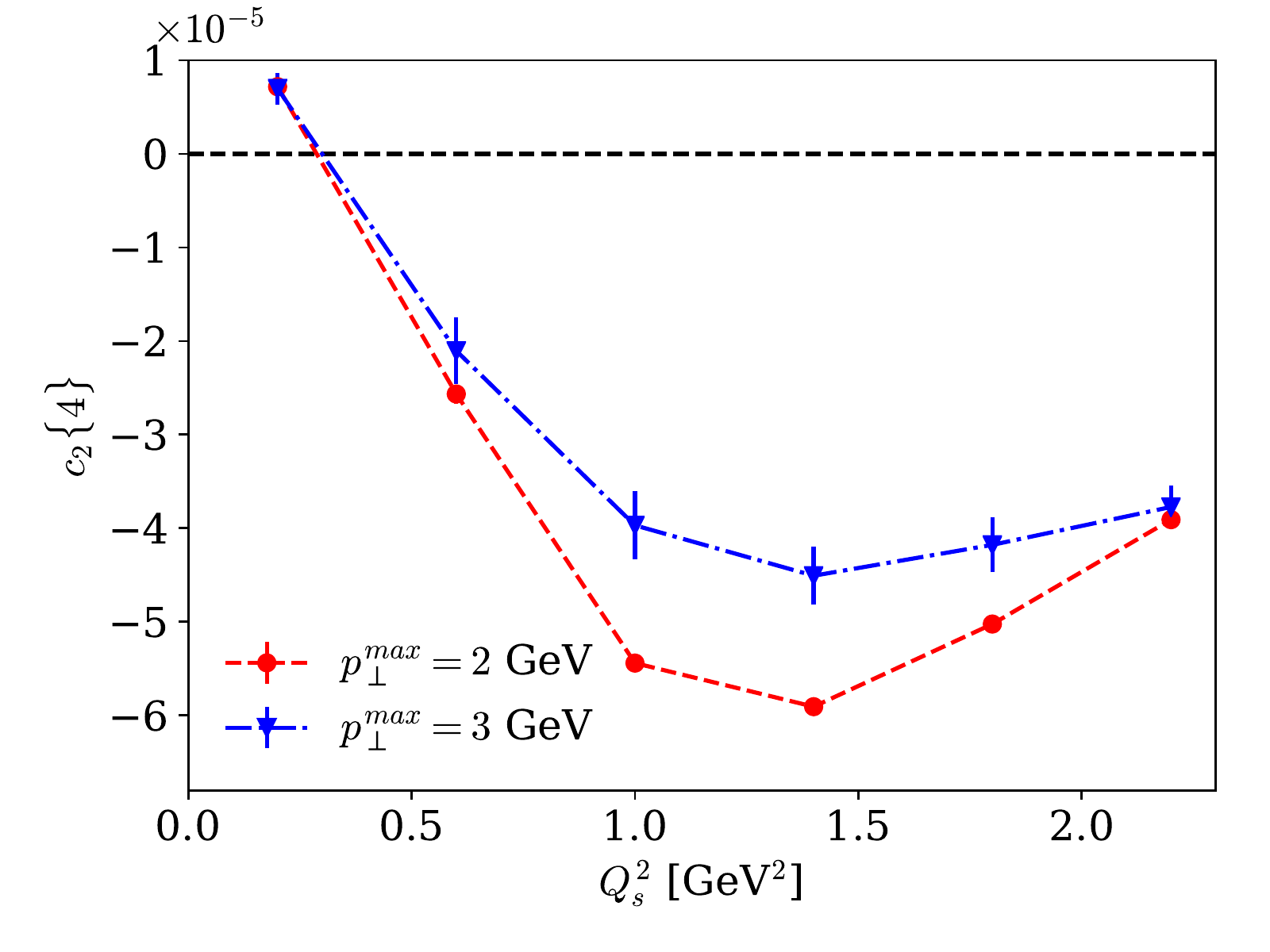}
\caption{$c_2\{4\}$ integrated to $p_\perp^{max}=2,3~\text{GeV}$ as a function of $Q_s^2$.}
\label{fig:c24_pT}
\end{minipage}
\end{figure}

In \Fig{fig:vn2_ptmax}, we show $v_n\{2\}$ as a function of $Q_s^2$ for Fourier harmonics $n=2,3,4,5$. For each harmonic, we studied the sensitivity of the result for multiple values of the maximum integrated transverse momentum ($\pmax$).  We observe that even for $\pmax=2$~GeV, the result is insensitive to the high momentum cutoff. Due to the small $x$ evolution of partons in the target, its saturation momentum $Q_s$ will increase  with the increasing center-of-mass energy of the collision. Since this is the only energy dependent variable in our framework, our result indicates that the $v_n\{2\}$ are only weakly dependent on the energy. We also observe that the $v_n\{2\}$ have a clear hierarchy with $n$, similar to what is seen in experiment~\cite{Aad:2014lta}.

The four-particle cumulant $c_2\{4\}$ is shown as a function of $Q_s^2$ in \Fig{fig:c24_pT}. We clearly see that by $Q_s^2=0.3~\text{GeV}^2$ that there is a change in the sign of the signal, from positive to negative values, resulting in a real $v_2\{4\}=(-c_2\{4\})^{1/4}$.  The magnitude of the signal only weakly depends on the maximum integrated momentum $\pmax$.  The relatively weak variation in the signal above $Q_s^2\sim 1$~GeV$^2$ is in qualitative agreement with the experimental findings on the center-of-mass energy dependence of the four-particle cumulant~\cite{Chatrchyan:2013nka,Abelev:2014mda,Khachatryan:2015waa,Belmont:2017lse,Aaboud:2017acw}.

As we noted previously, a positive definite value for $v_2\{4\}$ is natural in hydrodynamic models. However, we know of only two 3+1-dimensional numerical hydrodynamic computations of $v_2\{4\}$ in $pA$ collisions~\cite{Bozek:2012gr,Kozlov:2014fqa}. In the case of Ref.~\cite{Kozlov:2014fqa}, while some of the systematics of $pA$ collisions is reproduced, the model is unable to reproduce similar event-by-event systematics of flow in $AA$ collisions.

There also exist qualitative arguments for positive definite $v_2\{4\}$~\cite{Dumitru:2014yza} in an initial state ``color domain'' model~\cite{Kovner:2010xk,Dumitru:2014dra,Dumitru:2014vka,Dumitru:2015cfa,Skokov:2014tka}. Though this model provides an intuitive picture of how multiparticle azimuthal cumulants may be generated in an initial state framework, it is unclear how the oriented background color-electric fields are created from first principles~\cite{Lappi:2015vta}.

\begin{figure}
\includegraphics[width=0.45\textwidth]{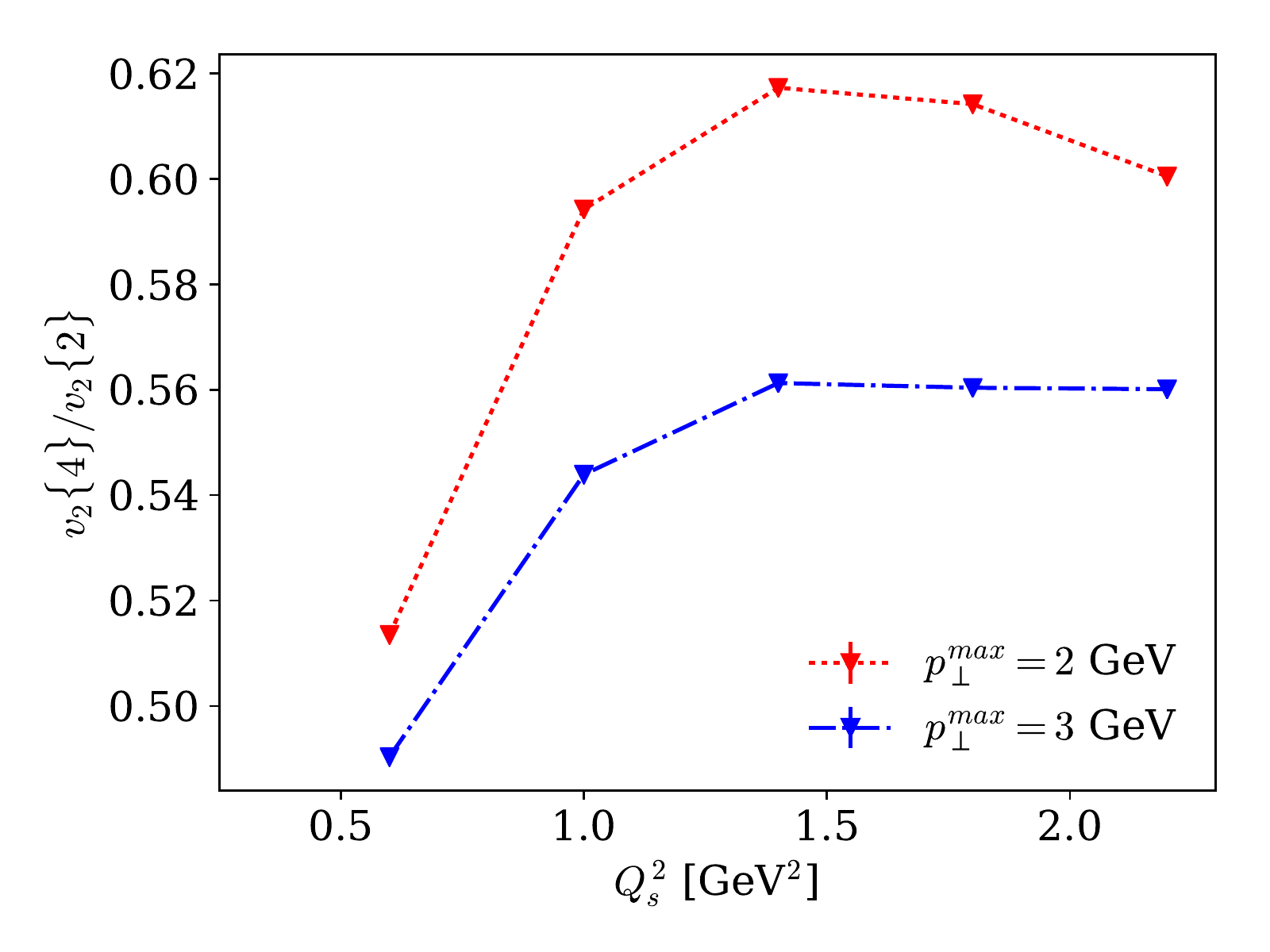}
\caption{The ratio of $v_2\{4\}/v_2\{2\}$ as a function of $Q_s^2$, for two values for $p_\perp$.}
\label{fig:v24overv22}
\end{figure}
In \Fig{fig:v24overv22}, we plot the ratio of the four- to two-particle integrated $v_2\{m\}$. For both values of $p_\perp^{max}$, this ratio rises with $Q_s^2$ and saturates above $Q_s^2\sim 1$~GeV$^2$.  The values obtained $v_2\{4\}/v_2\{2\}\sim 0.5$--$0.6$ are remarkably close to those measured by the CMS Collaboration that shows this quantity increasing with centrality from $\sim 0.675$ to $0.775$~\cite{Chatrchyan:2013nka,Giacalone:2017uqx}. A detailed study of this ratio, and other such ratios, and their centrality dependence, in various models of initial state spatial eccentricities, can be found in Ref.~\cite{Giacalone:2017uqx}.

\begin{figure}
\begin{subfigure}[t]{0.45\textwidth}
\includegraphics[width=\textwidth]{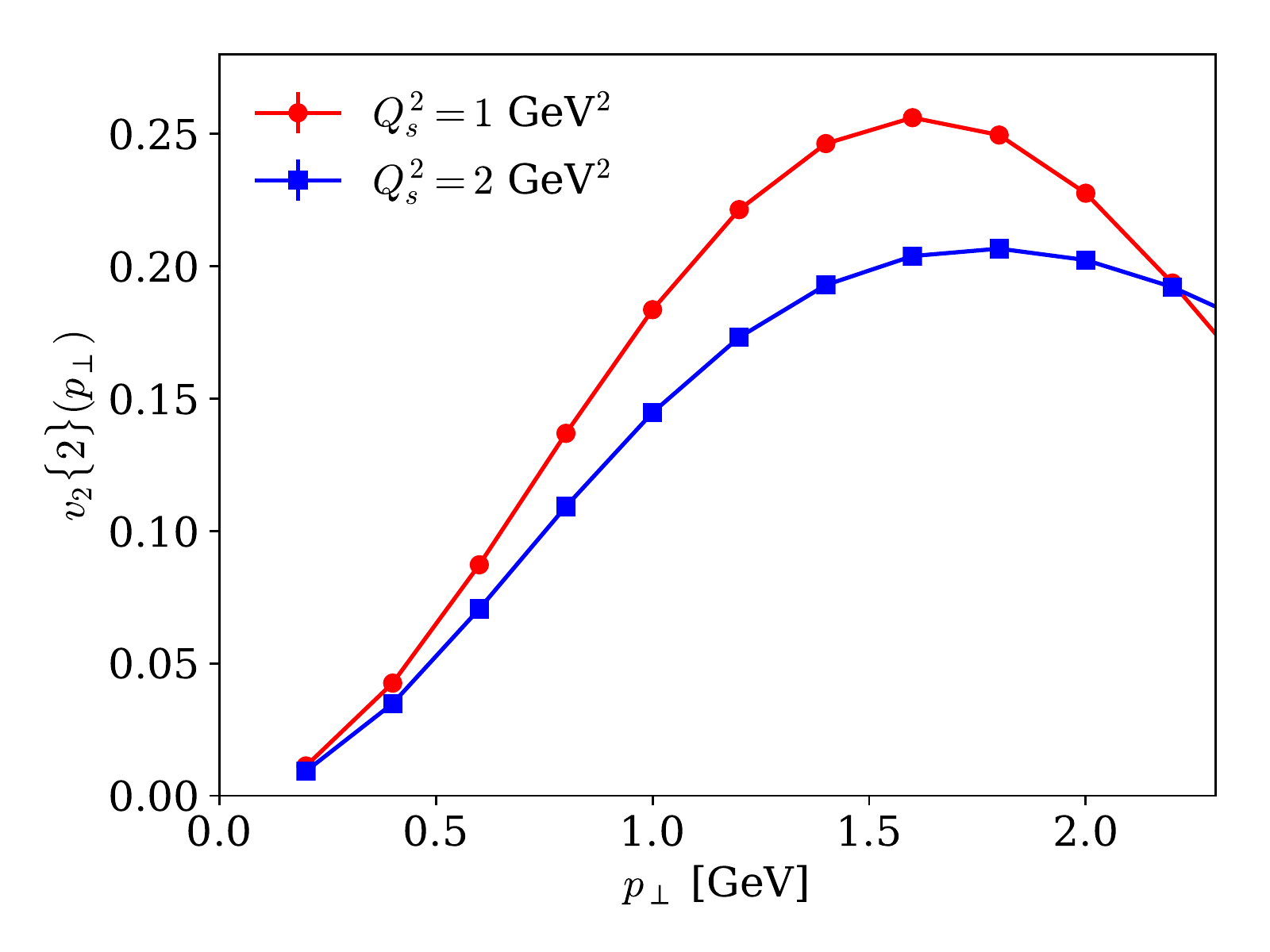}
\end{subfigure}
\begin{subfigure}[t]{0.45\textwidth}
\includegraphics[width=\textwidth]{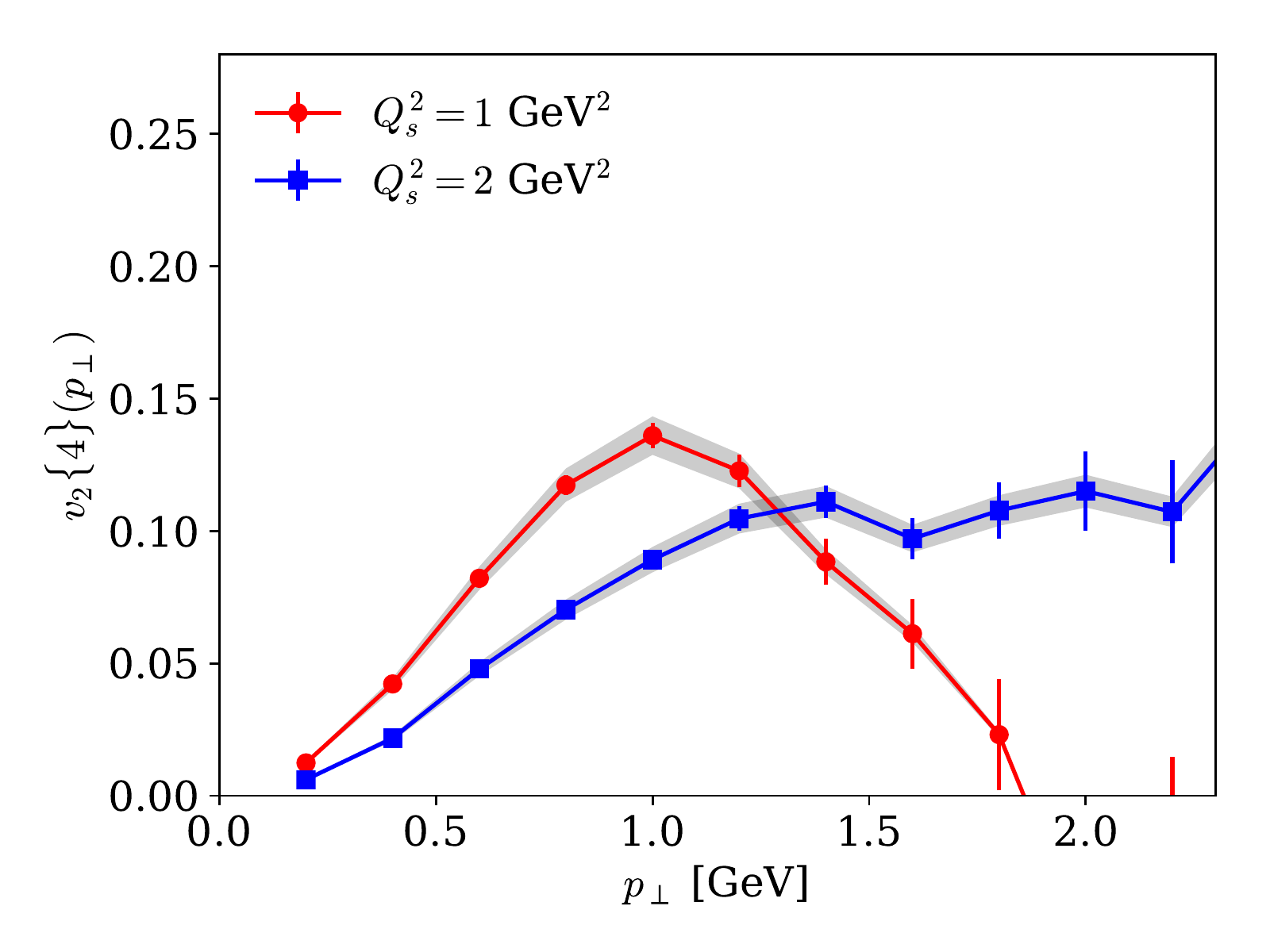}
\end{subfigure}
\caption{The Fourier harmonics $v_2\{2\}$ and $v_2\{4\}$ as a function of $p_\perp$ for two values of $Q_s^2$.}
\label{fig:d24_pt}
\end{figure}

We now study the multiparticle cumulants differentially in transverse momentum.  If we keep the transverse momentum of one particle in \Eq{eqn:kappa} fixed and integrate over the momenta of the remaining $m-1$ particles, the two- and four-particle differential Fourier harmonics are defined as~\cite{Chatrchyan:2013nka}
\begin{eqnarray}
\label{eqn:exp_v24}
v_2\{2\}(p_\perp)=\frac{d_2\{2\}(p_\perp)}{\left(c_2\{2\}\right)^{1/2}}\,,\qquad v_2\{4\}(p_\perp)=\frac{-d_2\{4\}(p_\perp)}{\left(-c_2\{4\}\right)^{3/4}}\, ,
\end{eqnarray}
where $d_n\{m\}$ are the differential analogs of $c_n\{m\}$.  Figure~\ref{fig:d24_pt} shows $v_2\{2\}$ (left) and $v_2\{4\}$ (right) as a function of $p_\perp$ for two representative values of the saturation scale $\Qs^2=1$ and $2$~GeV$^2$.  Note that while the results are differential in $p_\perp$ for one of the particles the remaining particles are integrated up to a $\pmax = 3$ GeV. We note that for $\Qs^2=2~\text{GeV}^2$ our results exhibit behavior similar to that seen in the experimental $p$-Pb data data~\cite{Chatrchyan:2013nka,Aad:2013fja}. We know of one other theory computation of $v_2\{4\}(p_\perp)$ in small systems~\cite{Iancu:2017fzn}.

Event-by-event fluctuations of $v_n$ with $v_{n^\prime}$ for $n\neq n^\prime$ can be captured by symmetric cumulants~\cite{Bilandzic:2013kga} defined as
\begin{eqnarray}
\text{SC}(n,n^\prime)&=&\langle e^{i(n\phi_1+n^\prime\phi_2-n\phi_3-n^\prime\phi_4)} \rangle-\langle e^{in(\phi_1-\phi_3)} \rangle \langle e^{in^\prime(\phi_2-\phi_4)} \rangle\,.
\label{eqn:symmetric_cumulant}
\end{eqnarray}
These mixed harmonic angular expectation values are defined analogously to those in~\Eq{eqn:kappa} and~\Eq{eqn:azimuth_corr}.
They were originally introduced in hydrodynamic frameworks as a measure of the nonlinear response of the initial geometry of the system~\cite{Bilandzic:2013kga}.  These have been studied mainly in the context of heavy-ion collisions~\cite{ALICE:2016kpq}. In these heavy-ion systems, the symmetric cumulants describe how correlations between the initial moments of the eccentricities, which are driven by the overlap geometry and thus the centrality of the collision, are carried to the final state.  An example of symmetric cumulants are the correlations between the $v_2$ and $v_3$ azimuthal harmonics, denoted by $\text{SC}(2,3)$. From geometrical considerations, there should be an anti-correlation between the initial ellipticity and the triangularity. When converted to correlations of final state momentum anisotropies, this results in a negative $\text{SC}(2,3)$. Studies of the symmetric cumulants $\text{SC}(2,3)$ and $\text{SC}(2,4)$ for heavy-ion collisions have been carried out in hydrodynamic simulations~\cite{Giacalone:2016afq,Zhu:2016puf} and in a hadronic transport model~\cite{Zhou:2015eya}.

\begin{figure}
\includegraphics[width=0.45\textwidth]{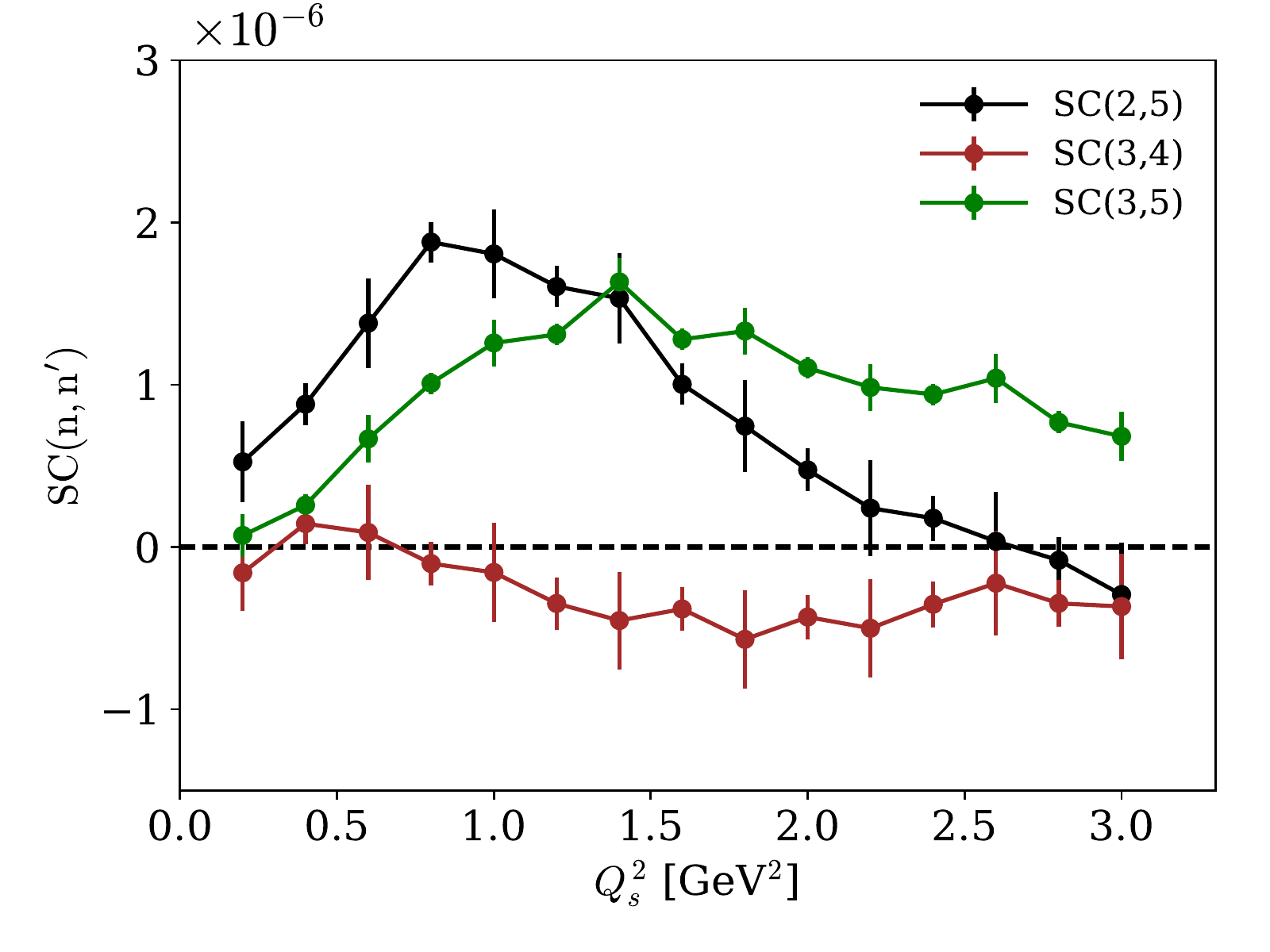}
\caption{Four-particle symmetric cumulants defined in \Eq{eqn:symmetric_cumulant} as a function of $\Qs^2$.}
\label{fig:SC}%
\end{figure}

However, in small systems, this picture should not hold.  The initial eccentricities are not believed to be strongly centrality driven, but instead are most likely in response to sub-nucleonic fluctuations~\cite{Welsh:2016siu}.  We will touch on the topic of sub-nucleon fluctuations in the next section. In our model, we include sub-nucleon scale fluctuations in sampling the positions of the quarks in the projectile given through our Gaussian Wigner function,~\Eq{eqn:wignergaussian}. We reported in \cite{Dusling:2017dqg} that our model produces the correct sign for $\text{SC}(2,3)$ and $\text{SC}(2,4)$ and is in qualitative agreement with the data~\cite{CMS:2017saf}. In \Fig{fig:SC}, we show  predictions for the behavior of $\text{SC}(2,5)$, $\text{SC}(3,4)$, and $\text{SC}(3,5)$. Results for these symmetric cumulants, for heavy-ion collisions alone, were shown at the Quark Matter 2017 conference~\cite{YZhou:QMtalk}. Our results for these cumulants agree qualitatively the results presented. We are unaware of any other theory predictions for these cumulants in light-heavy ion collisions. While not their designated purpose, symmetric cumulants in small systems may be an effective way to access information on initial state sub-nucleon  fluctuations.

\section{Detailed systematics of the results}
\label{sec:discussion}
\subsection{Role of the projectile}

It is expected that sub-nucleon scale fluctuations play an important role in small systems; hydrodynamic computations including such fluctuations have been performed for $pA$ collisions~\cite{Mantysaari:2017cni}. Thus it is also interesting to ask what similarities our model bears to a constituent quark model based picture. To mock up this effect, we introduce a hard distance cutoff (either minimum or maximum) between all quarks in the amplitude and similarly between all anti-quarks in the complex-conjugate amplitude. This is in addition to the Gaussian sampling of the quark positions from the Wigner function introduced in \Eq{eqn:wignergaussian}.

\begin{figure}
\includegraphics[width=0.45\textwidth]{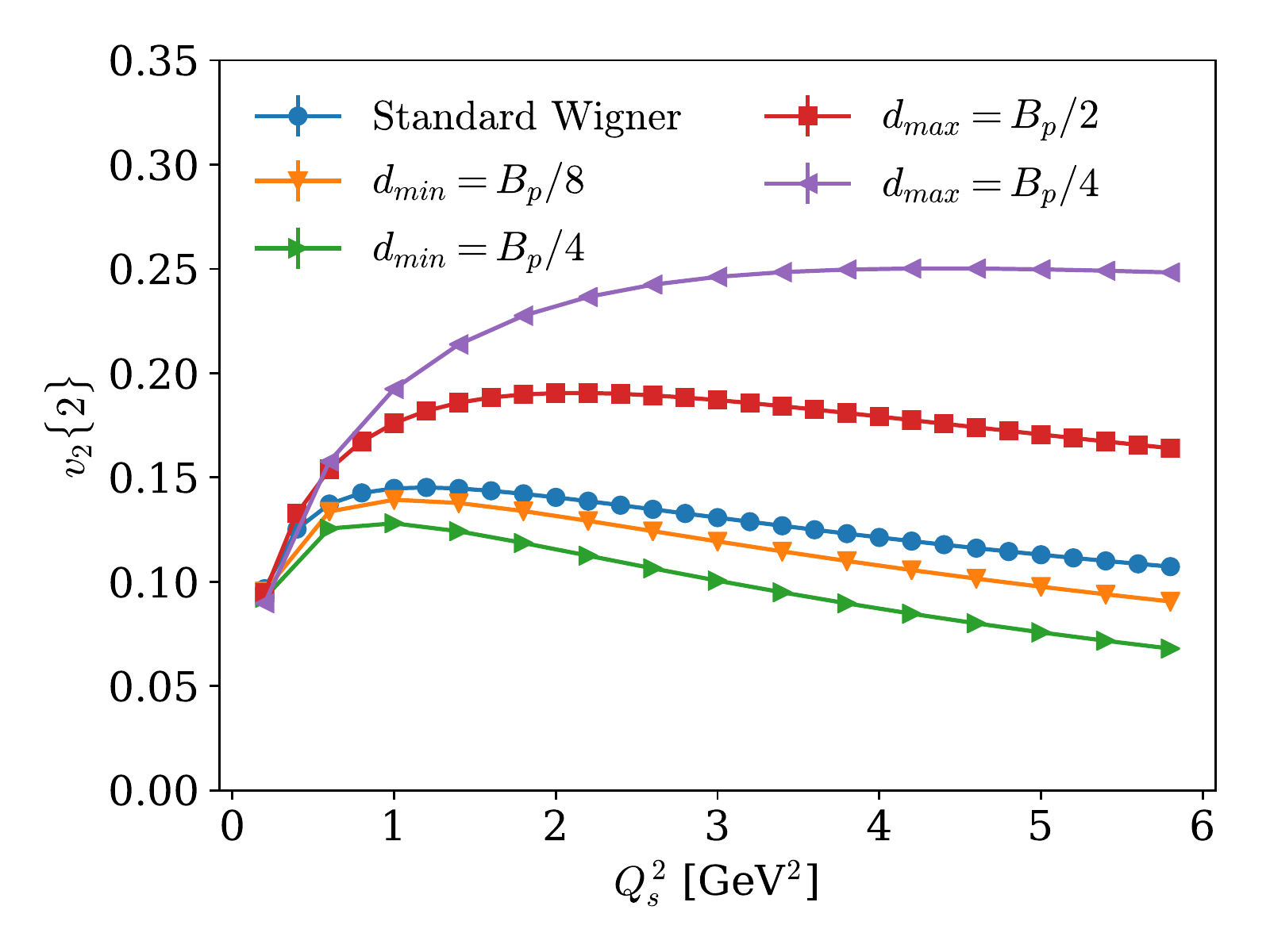}
\caption{$v_2\{2\}$ with minimal and maximal separations between quarks from the projectile. The ``standard Wigner" curve imposes no such constraint.  The orange and green curves show what happens when the quarks in the projectile are required to be separated by at least $d_{min}$.  The red and purple curves show the result when the quarks are required to be confined to a distance of $d_{max}$ from each other.}
\label{fig:v22_sep}
\end{figure}

The effect of such a cutoff on $v_2\{2\}$ is shown in \Fig{fig:v22_sep}. Starting with the standard Wigner function in \Eq{eqn:wignergaussian} as a reference, we see that by introducing a minimum distance criteria (separating the quarks by at least a distance of $d_{min}=\Bp/8$ or $\Bp/4$) the correlations decrease.  This is to be expected because forcing the transverse positions of the quarks to be farther away from each other ensures that they are less likely to interact with the same color domain in the target. We would then expect, on the same grounds, that if we confined the quarks to be at most a distance $d_{max}=\Bp/4$ or $\Bp/2$ apart, we would see an increase in the strength of the correlation.  This expectation is confirmed by the results shown in \Fig{fig:v22_sep}.

\subsection{$Q_s^2 \Bp$ dependence}
\label{sec:Qs2B_dep}

\begin{figure}
\includegraphics[width=0.45\textwidth]{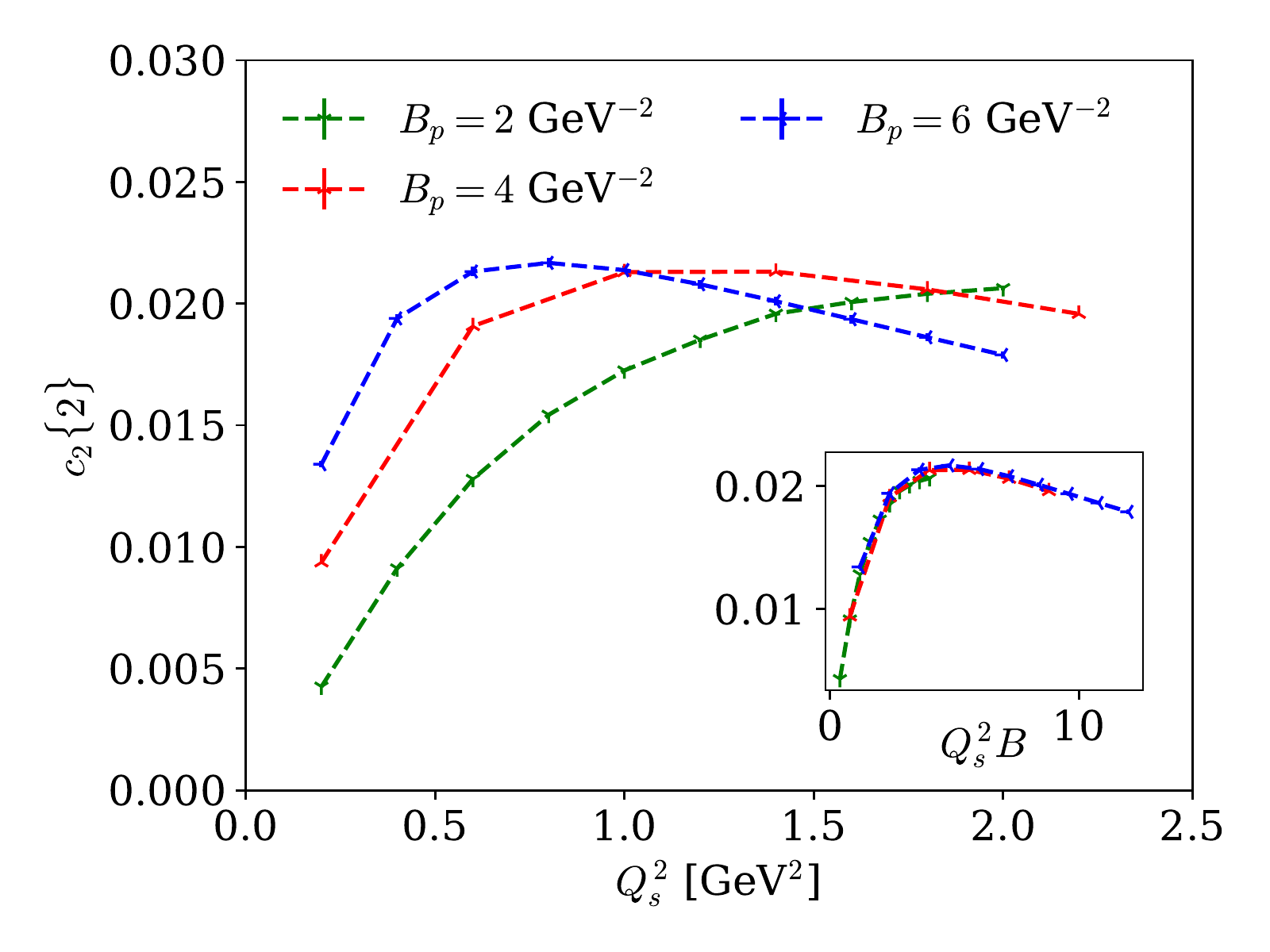}
\caption{The two-particle cumulant $c_2\{2\}$ plotted as a function of $Q_s^2$ for different values of $\Bp$, the mean transverse area of the projectile. Inset: Same data plotted as a function of the dimensionless ratio $Q_s^2\Bp$, showing the independence of $Q_s^2\Bp$.}
\label{fig:B_dep}
\end{figure}

In our model, the parameter $\Bp$ controls the mean transverse area of the projectile, and therefore the transverse overlap area of the scattering off the target. The scale $1/Q_s^2$ sets the scale for the size of color domains in the target. Therefore the dimensionless product $Q_s^2\Bp$ can be interpreted as the number of domains in the target that overlap the projectile. In fact, this dimensionless parameter controls the strength of all correlations. \Fig{fig:B_dep} shows $c_2\{2\}$ for three values of $\Bp$ for $\pmax=3~\text{GeV}$; note that $\Bp=4~\text{GeV}^{-2}$ is used elsewhere in this work.  The inset in \Fig{fig:B_dep} shows the same quantity plotted as a function of the dimensionless scale $Q_s^2\Bp$ demonstrating that all results fall onto a universal curve, as they must.

One might expect that the strength of the correlation should fall with the number of domains $1/(Q_s^2\Bp)$, which is a feature of independent cluster models.  However, a falloff that goes like $1/(Q_s^2\Bp)$ is not seen in the results presented above.  The reason is that there is another scale, $\pmax$, which controls the maximal momentum kick from the target to the probe. The inverse of $\pmax$ is therefore the smallest distance in the target resolved by the probe.
One can therefore construct two dimensionless combinations $Q_s^2 \Bp$ and $Q_s^2/({\pmax})^2$; the dependence of our results on the number of color domains also depends on what $Q_s^2/({\pmax})^2$ is.

\begin{figure}
\includegraphics[width=0.45\textwidth]{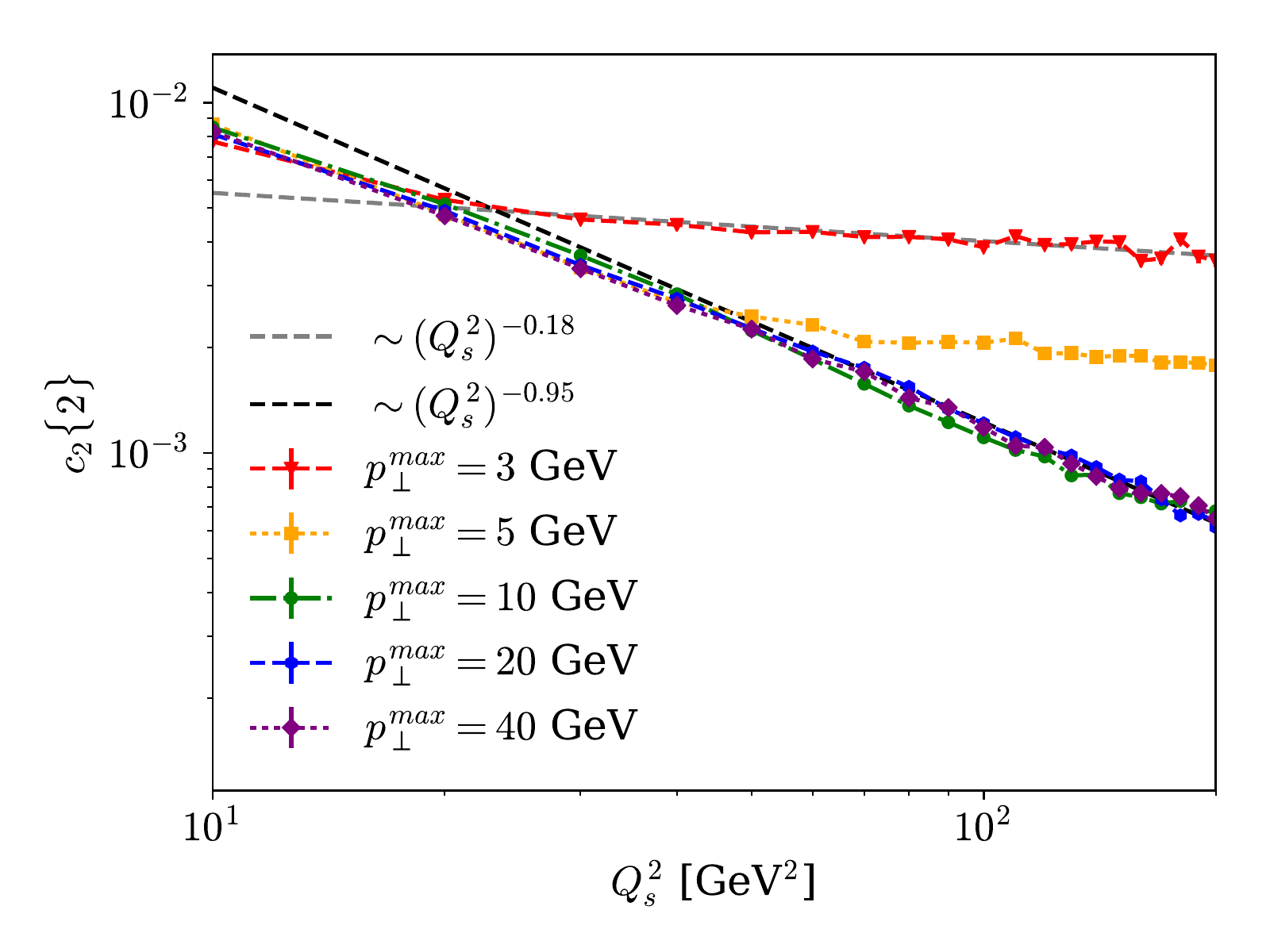}
\caption{Two-particle cumulant $c_2\{2\}$ for various $p_\perp^{max}$ for large values of $Q_s^2$ for fixed $\Bp=4~\text{GeV}^{-2}$. For $\pmax\leq \Qs$, only a weak dependence on $\Qs^2\Bp$ is seen. For larger $p_\perp^{max}$, we can see a falloff in the value of the cumulant that scales approximately with the number of color domains, as $\sim1/{Q_s^2\Bp}$.}
\label{fig:c22_largeQs}
\end{figure}

For $({\pmax})^2 \gtrsim Q_s^2$,  the probe resolves a transverse area in the target that is on the order or smaller than the size of a color domain.  Because the probe can resolve the structure within individual domains,  we expect to see a falloff in correlations to go approximately as $1/(Q_s^2\Bp)$.  \Fig{fig:c22_largeQs} shows $c_2\{2\}$ for $\pmax=10, 20, 40$ GeV, all of which satisfy the scaling form $(Q_s^2\Bp)^{-0.95}$ at large $\Qs^2$.
On the other hand, for $({\pmax})^2 \leq \Qs^2$, the probe only resolves transverse sizes larger than the typical domain size. For these smaller values of $\pmax$, increasing $\Qs^2\Bp$, and therefore the number of color domains, does not change the signal since the probe cannot resolve the change in the number of color domains. \Fig{fig:c22_largeQs} shows that for $\pmax=3,5$ GeV we see a rather modest falloff with the number of domains $(Q_s^2\Bp)^{-0.18}$.

Our results in \Fig{fig:c22_largeQs} suggest more generally that for small values of $\pmax$ relative to the saturation scale $\Qs$ azimuthal cumulants in initial state models are weakly dependent on the number of clusters. This independence of Fourier harmonics on the number of clusters has been interpreted previously as occurring due to the collective response of the system~\cite{Basar:2013hea}. While coherent multiple scattering may be collective, it is not a final state effect in $pA$ collisions; the interaction with the target is instantaneous and the scattered quarks do not subsequently rescatter.

\subsection{$N_c$ dependence}
\label{sec:Nc_dep}

\begin{figure}
\begin{subfigure}[t]{0.45\textwidth}
\includegraphics[width=\textwidth]{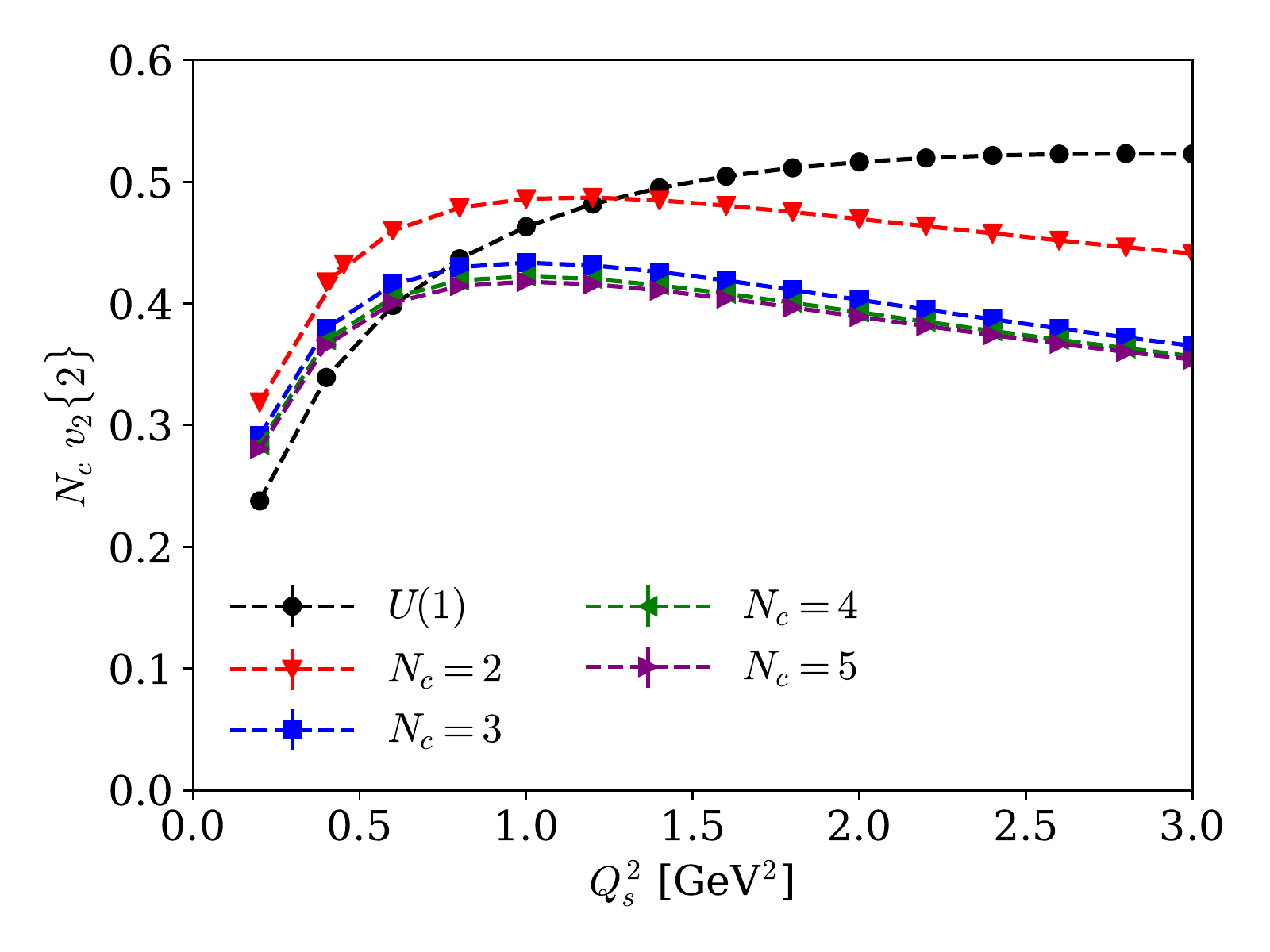}
\end{subfigure}
\begin{subfigure}[t]{0.45\textwidth}
\includegraphics[width=\textwidth]{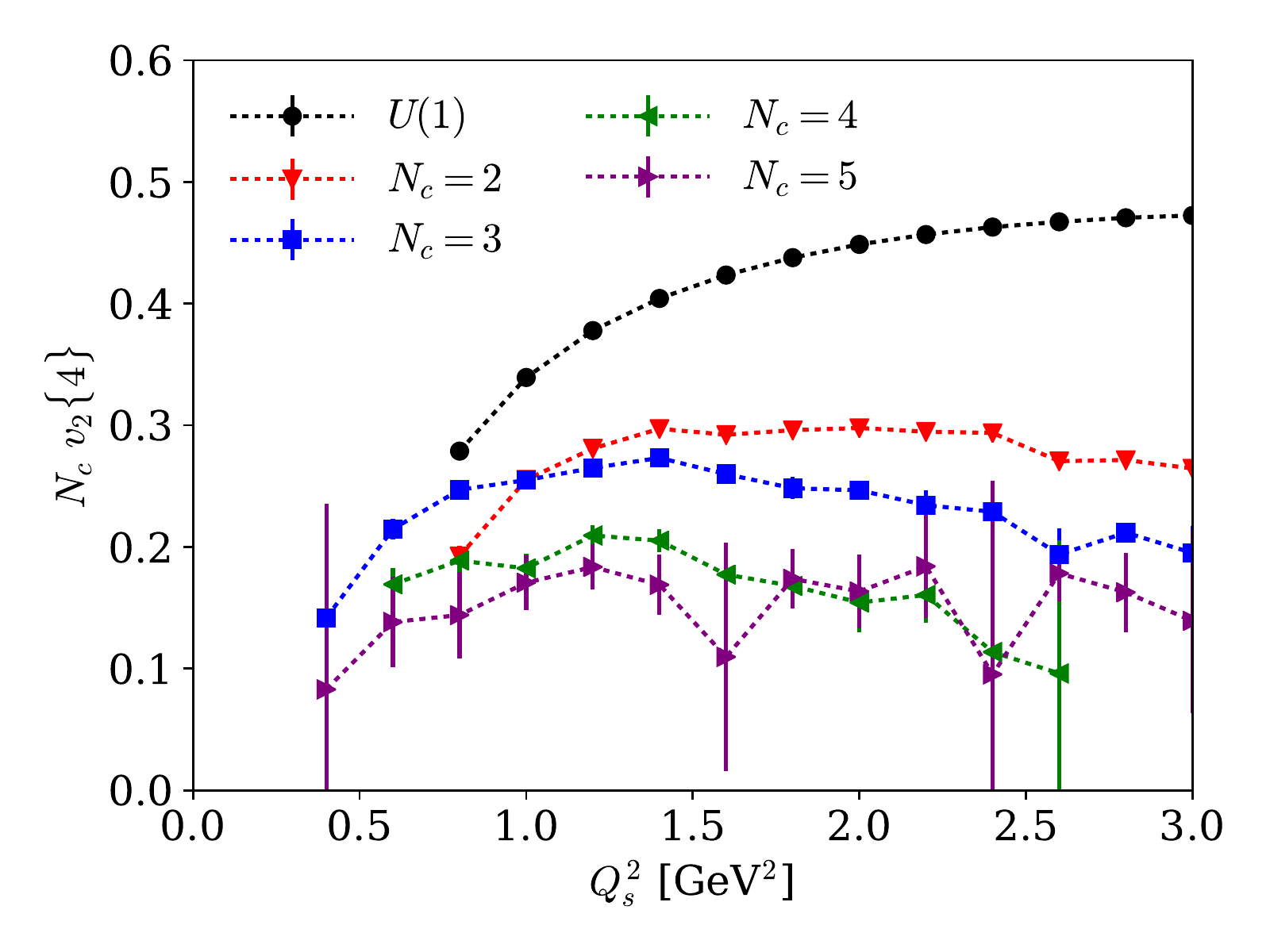}
\end{subfigure}
\caption{
The $N_c$ dependence of $v_2\{2\}$ (left panel) and $v_2\{4\}$ (right panel).
}
\label{fig:Nc}%
\end{figure}

In Ref.~\cite{Dusling:2017dqg}, we showed that an Abelianized version of our model demonstrates the systematics often attributed to collective behavior, $v_2\{4\}\approx v_2\{6\}\approx v_2\{8\}$.  Given this finding, it is natural (and of intrinsic interest) to determine the $N_c$ dependence of the two- and four-particle azimuthal correlations. The dependence of $v_2\{2\}(p_\perp)$ for $N_c=2,3$ was discussed previously in Ref.~\cite{Lappi:2015vta}. In the left panel of \Fig{fig:Nc}, we plot the dependence of integrated $v_2\{2\}$ (up to a $\pmax =2$ GeV) as a function of $\Qs^2$ for the Abelian ($N_c=1$) case and for $N_c=2-5$. For large $\Qs^2$, we observe a convergence of the results for $N_c\geq 3$. In the right panel of \Fig{fig:Nc}, we plot the $N_c$ dependence of $v_2\{4\}$ as a function of $\Qs^2$. When $v_2\{4\}$ is real and large, we expect the second term in \Eq{eqn:intrinsic-cumulants} for the four-particle cumulant to dominate. This should then give  $v_2\{4\} \sim 1/{N_c}$.  We see from \Fig{fig:Nc} that $N_c\,v_2\{4\}$ begins to converge for $N_c\geq 3$; however, due to limited statistics, the error bars are large.

\begin{figure}
\includegraphics[width=0.45\textwidth]{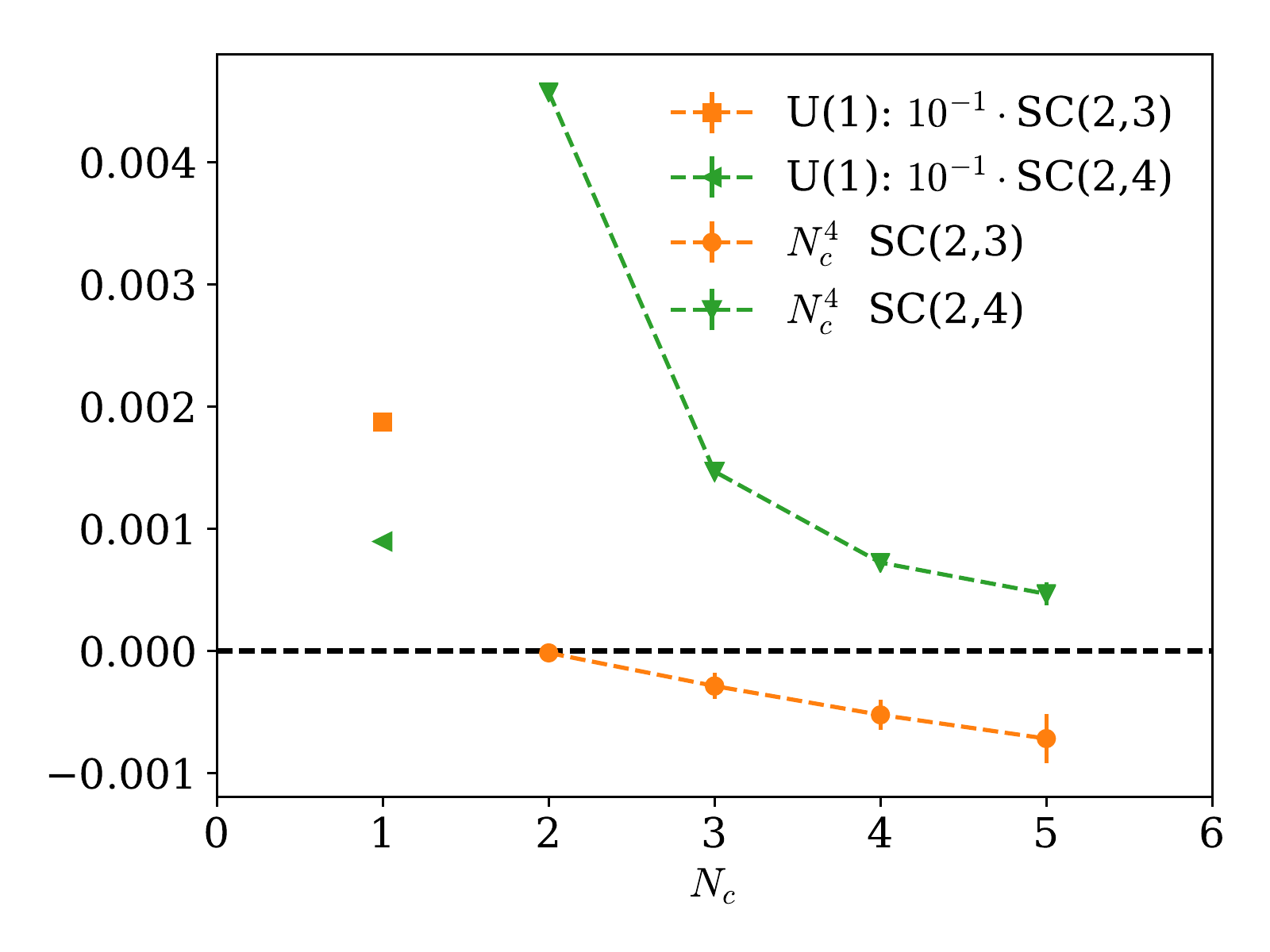}
\caption{$N_c$ dependence of the symmetric cumulants SC(2,3) and SC(2,4) for $Q_s^2=1~\text{GeV}^2$.}
\label{fig:gg_sc}
\end{figure}

We previously reported in Ref.~\cite{Dusling:2017dqg} on results for the symmetric cumulants for $\text{SC}(2,3)$ and $\text{SC}(2,4)$ which were in qualitative agreement with experimental results~\cite{CMS:2017saf}. In \Fig{fig:gg_sc}, we show the $N_c$ dependence of the symmetric cumulants. We find that these symmetric cumulants are extremely sensitive to $N_c$. (We have chosen here $\pmax=2$ GeV.) For the Abelian case, the result is an order of magnitude larger than the finite $N_c$ results. Further, $\text{SC}(2,3)$ is positive in the Abelian case, which is not observed in any experimental observations.  Interestingly, within the limited number of observables that we have studied, this appears to be the only place where the Abelian version of our model qualitatively differs from the non-Abelian results.

One may infer that there is something specific to the non-Abelian nature of coherent scattering that drives $\text{SC}(2,3)$ to become negative. Going to $N_c=2$, we find that $\text{SC}(2,3)$ is identically zero. This is not surprising, as for $N_c=2$, all odd harmonics are identically zero. This is analogous to the absence of odd harmonics for gluons scattering off a target~\cite{Schenke:2015aqa}. The underlying reason is that $SU(2)$ representations are real, regardless of whether they are in the fundamental or the in adjoint. For $SU(3)$, only the adjoint representation is real. Thus one expects qualitatively different results for even-odd cumulants for $N_c\geq 3$ relative to the Abelian model and for $N_c=2$.

\subsection{Comparison to Glasma graphs}
\label{sec:gg_comp}
To elucidate the mechanism responsible for the observed negativity of $c_2\{4\}$ for our model, we compare this result to that from the Glasma graph approximation. In this Glasma graph approximation, which is applicable for $p_\perp > Q_s$, the Wilson lines are expanded out to lowest nontrivial order in the gauge fields. The Glasma graph approximation is reviewed in the \hyperref[app:gg]{Appendix}, where we also compute the four-dipole correlation function in this approximation. This approximation was used previously to successfully describe two-particle correlations~\cite{Dumitru:2008wn,Dumitru:2010iy,Dusling:2012iga,Dusling:2012cg,Dusling:2012wy,Dusling:2013qoz,Dusling:2015rja}, especially near side ``ridge'' correlations.

\begin{figure}
\includegraphics[width=0.45\textwidth]{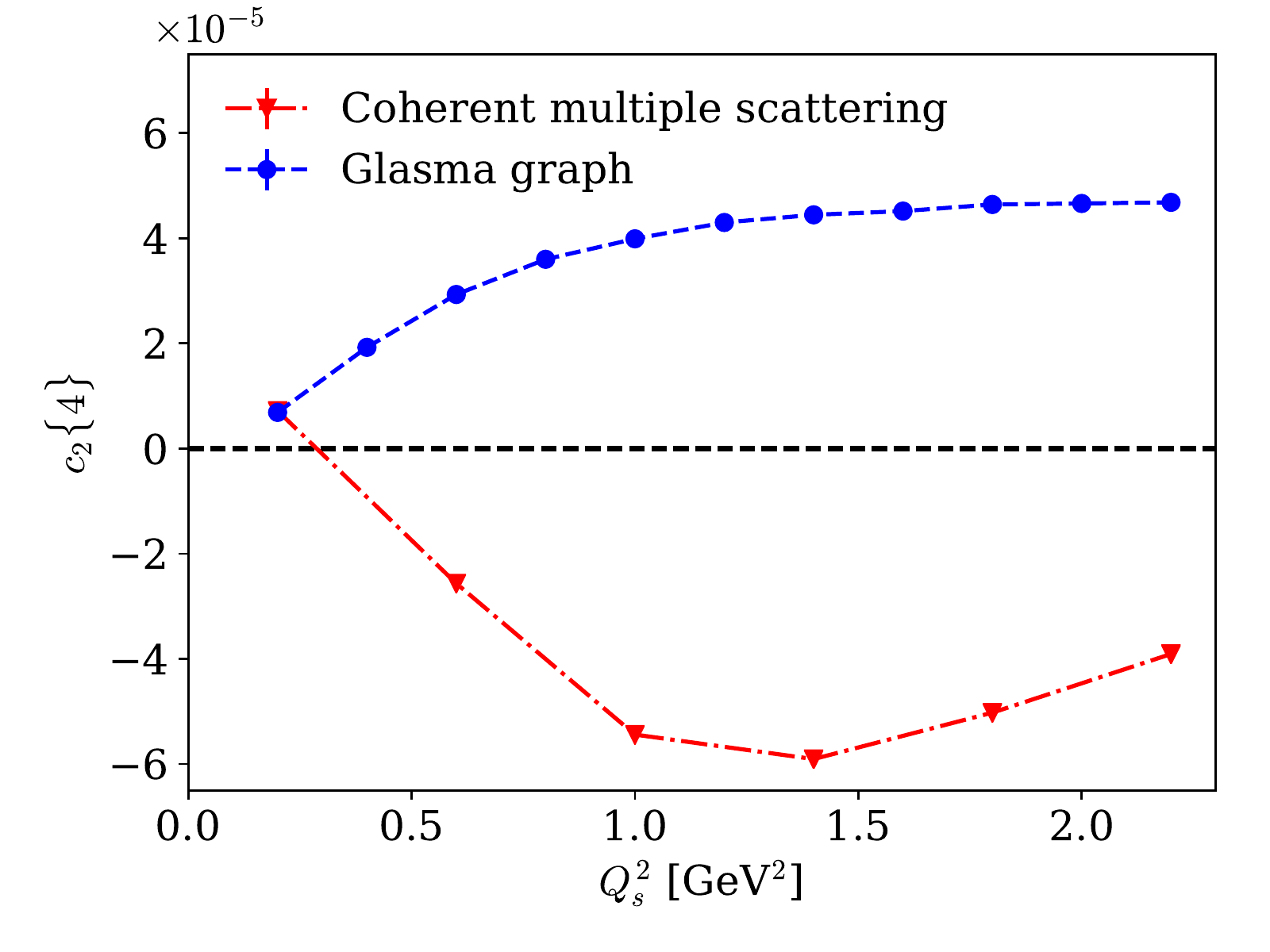}
\caption{Comparison of $c_2\{4\}$ in the model introduced in Sec.~\ref{sec:dipolecorr}, for coherent multiple scattering compared to the result, given in \Eq{eq:c24gg}, from the Glasma graph approximation.}
\label{fig:coherent-vs-glasma}
\end{figure}

\Fig{fig:coherent-vs-glasma} shows a comparison of $c_2\{4\}$ in the Glasma graph approximation to our result, which includes all order contributions from the Wilson lines. It is clear that coherent multiple scattering in the MV model computation for $c_2\{4\}$ drives it to become negative. In contrast, the Glasma graph approximation to this full result is always positive.

It is interesting to explore further the physics underlying this striking result. Intrinsic $n$-particle correlations in the Glasma graph approximation are large. Indeed, the strength of the correlated piece in this distribution relative to the disconnected product of $n$ single-particle distributions is the same for all $n$; this is close to that of an $n$-particle Bose distribution~\cite{Gelis:2009wh}. Our results suggest that coherent multiple scattering depletes these higher-point intrinsic correlations. In \Eq{eqn:intrinsic-cumulants} for instance, this would lead to the second term (the square of $c_2\{2\}$) to dominating over the first term, which comes from intrinsic four-particle correlations.

\subsection{Rapidity dependence}
\label{sec:rap_dep}

Before we conclude our discussion of features of the model, it is appropriate to discuss how rapidity correlations manifest themselves in this framework. Since long range rapidity correlations are an essential feature of the experimentally observed ridge-like correlations, it is important to determine whether such correlations are present in this framework. This is particularly so since the hybrid  formalism employed in analytical studies of such multiparticle correlations~\cite{Dumitru:2002qt,Kovner:2012jm,Kovchegov:2012nd}  is valid in the forward rapidity region. More precisely, $x$ in the projectile should be relatively large, with typical values for {\em large-$x$} taken to be $x\geq 0.01$. However this does not imply that the resulting correlations are short-range in rapidity. We will show this explicitly by reintroducing rapidity dependence into the single-particle and multi-particle distributions.

We consider an eikonal quark in the projectile traveling in the $z^+$ direction with initial momentum $k^\mu=(k^+,0,{\bf k_\perp}={\bf 0})$ and final momentum $p^\mu=(p^+,0,{\bf p_\perp})$ after its scattering off the target. In the hybrid framework, the differential multiplicity of the scattered quark with the final state of momentum ${\bf p}\equiv(p^+,{\bf p_\perp}$) is given as
\begin{equation}
\frac{dN^{qA\to q+X}}{d^3{\bf p}}\equiv\frac{dN^{qA\to q+X}}{dp^+ d^2{\bf p_\perp}}=\delta(p^+-k^+)\frac{dN^{qA\to q+X}}{d^2{\bf p_\perp}}\,.
\label{eqn:rap1}
\end{equation}
The above result can be worked out following the formalism of Ref.~\cite{Dumitru:2002qt}. (Note though that in Ref.~\cite{Dumitru:2002qt}, the quark is traveling in the $z^-$ direction.) The expression for $dN/d^2{\bf p_\perp}$ is given by \Eq{eqn:singleinclusive}, where averaging over the target we previously defined is implicitly assumed.  Therefore, for single-quark scattering the distribution is a delta function in $\delta(p^+-k^+)$, which is simply a consequence of the eikonal approximation.

The single inclusive distribution of quarks produced in $pA$ collisions is obtained by convoluting the above expression with the quark parton distribution, which represents the probability of finding a quark in the proton wavefunction:
\begin{equation}
\frac{dN^{pA\to q+X}}{d^3{\bf p}}=\int\,dx_q f(x_q) \frac{dN^{qA\to q+X}}{d^3{\bf p}}\,.
\label{eqn:rap2}
\end{equation}
To obtain the single inclusive distribution of hadrons, this expression has to further convoluted with a fragmentation function.  This will quantitatively modify the rapidity dependence but will not modify it qualitatively. We will therefore not further consider this point.

The longitudinal momentum carried by the initial quark is $k^+=\frac{\sqrt{s}}{\sqrt{2}} x_q$, where $x_q$ is the quark's momentum fraction.  Likewise, the longitudinal momentum of the final-state quark can be written in terms of its rapidity as $p^+=\frac{p_\perp}{\sqrt{2}}e^y$. Substituting \Eq{eqn:rap1} into \Eq{eqn:rap2}, we then obtain
\begin{equation}
\frac{dN^{pA\to q+X}}{dy d^2{\bf p_\perp}}=x_q' f(x_q') \frac{dN^{qA\to q+X}}{d^2{\bf p}_\perp}\,,
\end{equation}
where $x_q'$ has been set by the $\delta$-function to be
\begin{equation}
x_q'=\frac{p_\perp}{\sqrt{s}}e^y\,.
\end{equation}

\begin{figure}
\begin{subfigure}[t]{0.45\textwidth}
\includegraphics[width=\textwidth]{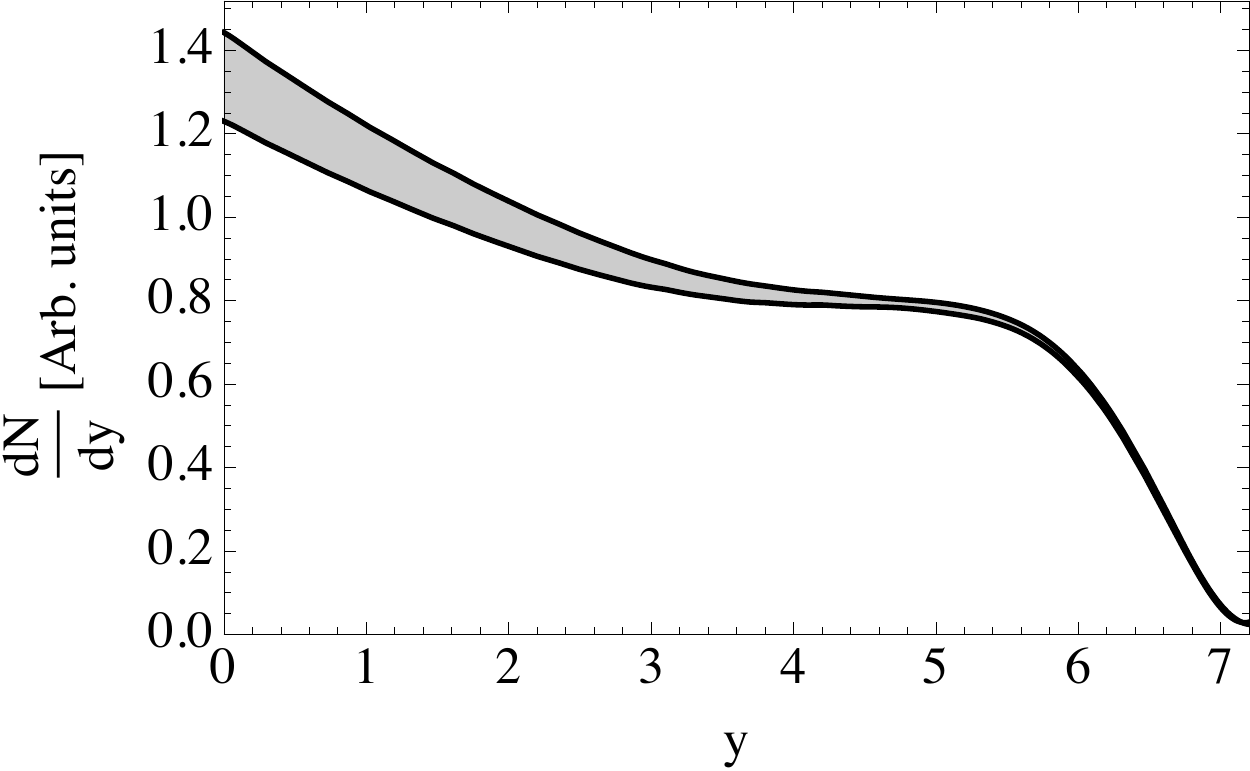}
\caption{Single particle distribution $dN/dy$ in arbitrary units as a function of rapidity for $p_\perp=3$~GeV and $\sqrt{s}=5.02$~TeV.}
\label{fig:dndy}
\end{subfigure}
\hspace{0.2cm}
\begin{subfigure}[t]{0.45\textwidth}
\includegraphics[width=\textwidth]{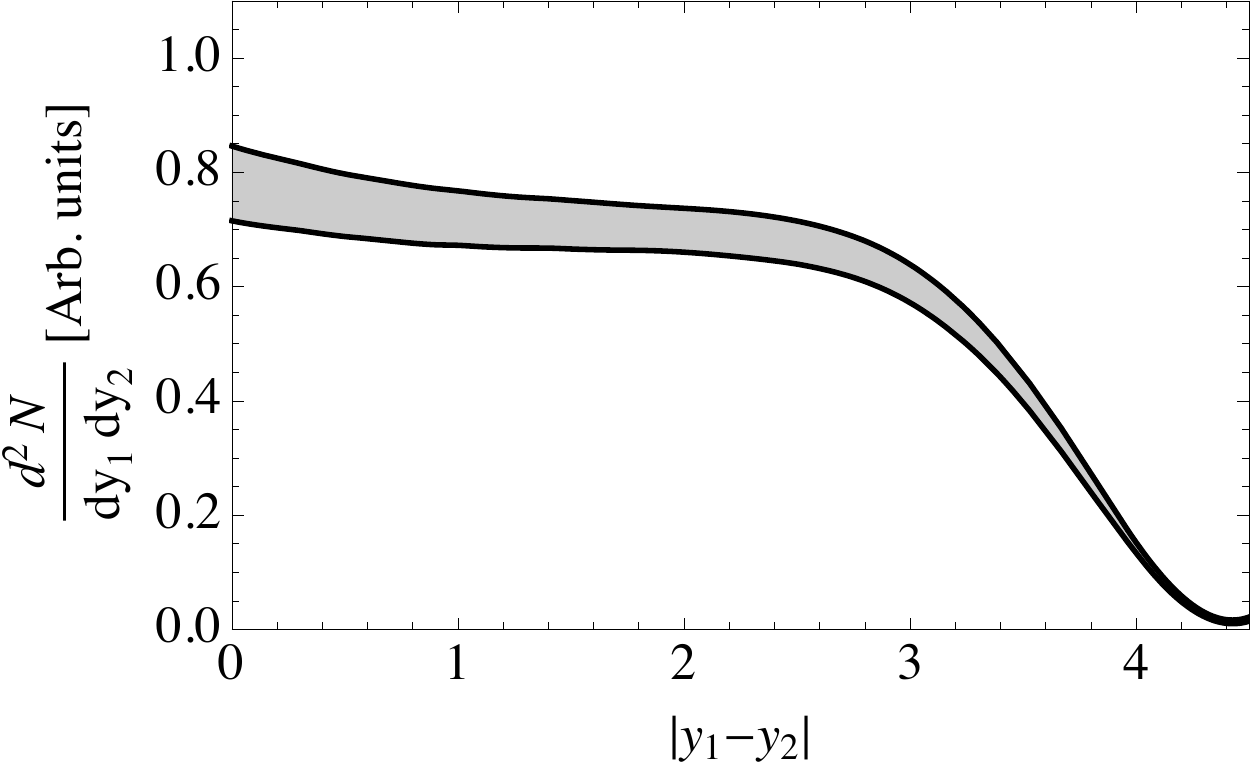}
\caption{Two-particle distribution $d^2N/dy_1dy_2$ in arbitrary units as a function of $\Delta y=\vert y_1-y_2\vert$ for $y_1\simeq 2.8$, which corresponds to $x_{q1}'=0.01$ in the projectile.}
\label{fig:d2ndy}
\end{subfigure}
\caption{}
\end{figure}

In \Fig{fig:dndy}, we show the single-particle inclusive quark distribution using the NNPDF NLO singlet PDF at $Q^2=9~$GeV$^2$~\cite{Ball:2014uwa}.  For $p_\perp=3$ GeV and $\sqrt{s}$ = 5.02~TeV, a value of $x_q=0.01$ corresponds to a rapidity of $y\simeq 2.8$. Thus, while strictly speaking, our approach is valid for $y\geq 2.8$, we can see that the particle production does extend over a wide range in rapidity and is not peaked in the forward direction.

This can be extended to multiparticle production in the same fashion. For instance, for two particles, we would obtain
\begin{equation}
\frac{d^2N^{pA\to q+X}}{dy_1 d^2{\bf p_{1,\perp}}dy_2 d^2{\bf p_{2,\perp}}}=x_{q1}' f(x_{q1}')x_{q2}'f(x_{q2}') \frac{d^2N^{qA\to q+X}}{d^2{\bf p_{1,\perp}}d^2{\bf p_{2,\perp}}}\,.
\label{eq:rapfact}
\end{equation}
This two-particle distribution is shown in \Fig{fig:d2ndy}. The rapidity of the first quark is fixed at the edge of where the hybrid approach works.  We see that the correlation persists out to a $\Delta y$ of 3 to 4 units as the rapidity of the second quark is varied.

We conclude by pointing out that when computing $v_n\{m\}$ we are taking ratios of the momentum integrated $m$-particle distributions, where the numerator is also convoluted with a cosine function. The factorized form of the rapidity dependence in \Eq{eq:rapfact} is then suggestive that the resulting $v_n\{m\}$ will be weakly dependent, or perhaps even constant, as a function of rapidity.

\section{Conclusions}
\label{sec:conclusion}

In this work, we elaborated significantly on a parton model framework for multiparticle correlations that was first presented in Ref.~\cite{Dusling:2017dqg}. In this model, collinear partons that are localized in a transverse area $\Bp$ in the projectile, scatter off color domains of size $1/\Qs^2$ in the target. The building blocks in this framework are dipole correlators. For a quark projectile, these correspond to a color trace over a path ordered lightlike Wilson line of target fields at a given transverse spatial position in the amplitude multiplied by its adjoint in the complex-conjugate amplitude, normalized by the number of colors, $N_c$. The azimuthal cumulants of $n$-particles are proportional to the expectation value over the product of $n$-dipole correlators. We discussed at length the procedure to compute the expectation value of four-dipole correlators. Gluon exchanges among the dipoles generate distinct quadrupole, sextupole, and octupole topologies and the permutations of their spatial positions, generating a 24 by 24 matrix, that can be exponentiated to determine the expectation value of four-dipole correlators. Our procedure can be extended straightforwardly to compute expectation values of products of an arbitrary number of dipole operators.

We presented results for $v_n\{2\}$ as a function of $\Qs^2$, demonstrating a clear hierarchy in the $n=2,3,4,5$ harmonics. These results are only weakly dependent on $\pmax$ upper limit in the integrals. All the harmonics show only a weak dependence on $\Qs^2$. Since $\Qs^2$ in the CGC framework increases with increasing energy, and centrality, our results are only weakly dependent on these. Note further that our results, by construction, are independent of the multiplicity.
These results for $v_n\{2\}$ exhibit the qualitative features of the data seen in light-heavy ion collisions at RHIC and the LHC.

We next presented results for $c_2\{4\}$ as a function of $\Qs^2$ for two different values of $\pmax$. For both these values, $c_2\{4\}$ changes sign around $\Qs^2=0.3$ GeV$^2$ and becomes increasingly negative before appearing to saturate. For large $\Qs^2$,  the results are sensitive to $\pmax$. The negative value of $c_2\{4\}$ corresponds to a real $v_2\{4\}$. We computed the ratio of $v_2\{4\}/v_2\{2\}$. In hydrodynamic models, such ratios are sensitive to the initial geometry in the system, motivating experimental extractions of the same. The values we obtained are about 10\% lower than the data, which at present have significant error bars.

The dependence of $v_2\{2\}$ and $v_2\{4\}$ as a function of $p_\perp$ increases with $p_\perp$ before saturating and turning over. For both quantities, this saturation occurs later with increasing $\Qs^2$. In particular, for $v_2\{4\}$, we observed for $\Qs^2=2$ GeV$^2$ that it is quite flat for the $p_\perp$ range between 1 and 2 GeV. These features of our results are also qualitatively similar to data on small-particle systems. To the best of our knowledge, no computations exist in other models for $v_2\{4\}(p_\perp)$.

Symmetric cumulants $\text{SC}(n,n^\prime)$, which measure the correlation of $n$th Fourier harmonics with $n^\prime$ Fourier harmonics, were constructed to understand the nonlinear hydrodynamic response of the system to correlations in the initial spatial geometry. We studied these in our initial state framework as a function of $\Qs^2$. We showed in Ref.~\cite{Dusling:2017dqg} that $\text{SC}(2,3)$ and $\text{SC}(2,4)$ computed are in qualitative agreement with the data presented for heavy-ion collisions and in light-heavy ion collisions. Here, we make predictions for the $\text{SC}(3,4)$, $\text{SC}(2,5)$, and $\text{SC}(3,5)$ cumulants.

We examined closely the dependence of our results on $Q_s^2 \Bp$ and $Q_s^2/({\pmax})^2$, the two dimensionless parameters in our model. The former corresponds to the number of color domains in the target that are encountered by the projectile. The latter corresponds to the resolution of the partons in the projectile to the structure of the color domains. Interestingly, we find that for larger values of $Q_s^2/({\pmax})^2$, the two-particle cumulants are only weakly dependent on the number of color domains. In contrast, for smaller values of $Q_s^2/({\pmax})^2$, we find that the cumulant behaves approximately as $1/(\Qs^2\Bp)$, as would be anticipated in an independent cluster model. Our results suggest therefore that the $\pmax$ considered is important in any interpretation of the data that may be construed as satisfying or violating an independent cluster model.

We studied next the dependence of our results on $N_c$. In \cite{Dusling:2017dqg}, we showed that the Abelianized treatment of our model reproduced the pattern of $v_2\{2\} > v_2\{4\} \approx v_2\{6\} \approx v_2\{8\}$ seen in the data on $pA$ collisions at the LHC. We studied further the $N_c$ dependence of $v_2\{2\}$ and $v_2\{4\}$ and demonstrated that the both behave as $1/N_c$ for $N_c\geq 3$. There is therefore no ordering in $N_c$ among $m$-particle cumulants. Practically, it means that $N_c$ suppressed topologies in products of lightlike Wilson lines must be included in such computations.

Unlike the azimuthal cumulants $c_2\{4\}$, the symmetric cumulants $\text{SC}(n,n^\prime)$ (where $n$ and $n^\prime$ denote distinct Fourier harmonics) have a qualitatively different behavior in the Abelian formulation of the model relative to those for the $N_c=3$ case we studied previously in Ref.~\cite{Dusling:2017dqg}. This qualitative difference is not unique to $N_c=1$. It is also seen for $N_c=2$. In this latter case, the odd harmonics are strictly zero; hence, the corresponding symmetric cumulants are also zero. As we discussed, the underlying reason is that for $SU(2)$ both fundamental and adjoint representations are real. This is responsible for two-particle correlations being symmetric about relative azimuthal angles of zero and $\pi$.

To obtain deeper insight into our results, we examined our results within the Glasma graph approximation to our framework. This approximation corresponds to expanding out the Wilson lines in the dipole correlators to lowest nontrivial order. Physically, it corresponds to each quark line interacting at most with two gluons, either in the amplitude, or the complex-conjugate amplitude, or across the cut separating the two. It is justified when $\Qs^2 \ll p_\perp^2$. Remarkably, we find that our results for $c_2\{4\}$ are qualitatively different when we include coherent multiple scattering ($\Qs^2/p_\perp^2$ to all orders) as opposed to the Glasma graph approximation. In the former case, one obtains a real and positive $v_2\{4\}$; in the latter case, one does not. Therefore, the absence of $v_n\{m\}$ multiparticle correlations in previous Glasma graph treatments is an artifact of the approximation and not a genuine feature of initial state frameworks.

We noted that multiparticle correlations are quite strong in the Glasma graph approximation, and are close to those of an $m$-particle Bose distribution. Indeed, this may be the reason why one does not see signatures of ``collectivity", as defined by the $v_n\{m\}$ Fourier harmonics. These assume that the mean and variance dominate the cumulant distribution. Our results suggest that coherent multiple scattering dilutes the contributions from the higher cumulants relative to the mean and the variance, thereby generating the aforementioned signatures of collectivity. In our model, however, the coherent scattering is virtually instantaneous. It takes place on a time scale corresponding to the time it takes partons to cross a Lorentz contracted nucleus. Further, the partons that scatter off the common color field do not rescatter.

The origin of this putative signature of collectivity therefore has little to do with hydrodynamics {\em per se}. However, our results do not exclude the possible presence of final state interactions, or even hydrodynamics, in the data on light-heavy ion systems. They do provide a clear and simple counterexample to interpretations of these signatures as being of unique origin. Such signatures of collectivity must also be consistent with other signatures of collectivity. In the larger heavy-ion collision systems, jet quenching is seen very clearly, and is an independently robust measure of final state interactions.

It is interesting to consider how this model can be developed further. Gluon degrees of freedom, which are not apparent at lower energies, become important when hadrons are boosted to higher energies. As is well known, a bremsstrahlung cascade develops, which then has a {\em shock wave} interaction with the nucleus. The partons in the cascade subsequently fragment to hadrons. This picture is implicit in the CGC+PYTHIA model of so-called dense-dense collisions of small systems~\cite{Schenke:2016lrs} that combines Yang-Mills dynamics of gluons~\cite{Schenke:2015aqa} with Lund fragmentation~\cite{Andersson:1983ia}. Because event-by-event simulations are essential to compute multiparticle cumulants, such computations are computationally intensive. Our work suggests the need for further developments in this direction.

\section{Acknowledgements}
We would like to thank Jiangyong Jia, Tuomas Lappi, Aleksas Mazeliauskas, Larry McLerran, Jean-Yves Ollitrault, Jean-Fran\c{c}ois Paquet, Bj\"{o}rn Schenke, S\"{o}ren Schlichting, Juergen Schukraft, Chun Shen, Vladimir Skokov, and Prithwish Tribedy for useful discussions. This material is based on work supported by the U.S. Department of Energy, Office of Science, Office of Nuclear Physics, under Contracts No. DE-SC0012704 (M.M. and R.V.) and No. DE-FG02-88ER40388 (M.M.). M.M. would also like to thank the BEST Collaboration for support. This research used resources of the National Energy Research Scientific Computing Center, a DOE Office of Science User Facility supported by the Office of Science of the U.S. Department of Energy under Contract No. DE-AC02-05CH11231 and the LIRED computing system at the Institute for Advanced Computational Science at Stony Brook University.
%
%
\appendix
\setcounter{secnumdepth}{0}
\section{Appendix: Glasma graph approximation}
\label{app:gg}
In this section, we show how our present calculation fundamentally differs from the so-called ``Glasma graph" result~\cite{Dumitru:2008wn,Dumitru:2010iy,Dusling:2012iga,Dusling:2012cg,Dusling:2012wy,Dusling:2013qoz,Dusling:2015rja,Ozonder:2014sra,Ozonder:2016xqn}. The Glasma graph approximation is constructed by considering all possible two-gluon exchanges between quarks comprising the projectile and the target nucleus under the assumption of Gaussian statistics. The expectation value of $n$-Wilson lines in the Glasma graph approximation can be evaluated by expanding the path ordered exponential of the Wilson line to order $n$ in the coupling constant.  The resulting expectation values of gauge fields are then evaluated by re-expressing higher order expectation values as a product of two-point functions.  This procedure was followed in Ref.~\cite{Lappi:2015vta} in order to evaluate the dipole-dipole correlator in the Glasma graph approximation.  Here, we extend the derivation and results of Ref.~\cite{Lappi:2015vta} to higher order correlators but take a more diagrammatic approach.

\begin{figure}[ht]
\scalebox{.5}{\includegraphics[]{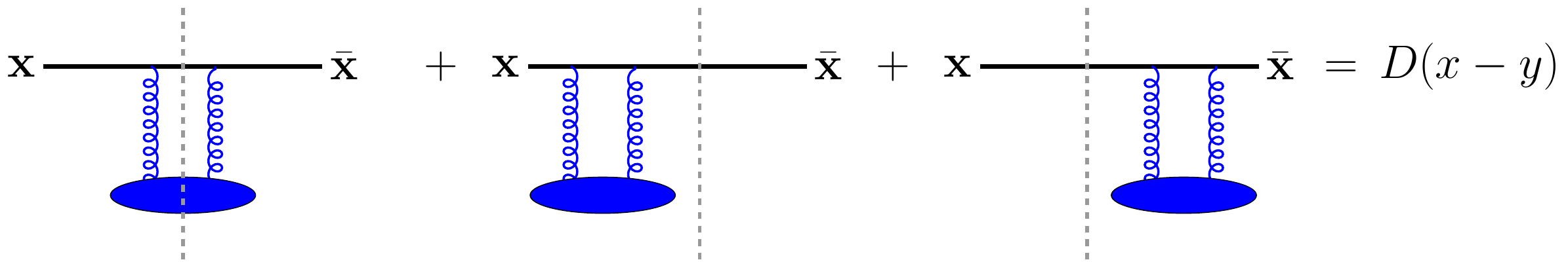}}
\caption{Single-quark multiple-scattering off the target nucleus.}
\label{fig:Tsingle}
\end{figure}

Diagrammatically, the Glasma graph approximation amounts to replacing each two-gluon exchange with the expectation value of a single dipole operator.  For example, for single-quark scattering we find the relation of \Fig{fig:Tsingle}.
While the tadpole terms are explicitly shown in the single scattering case in~\Fig{fig:Tsingle}, for higher-point functions diagrams containing two gluons on the same quark are power suppressed by either $k_\perp/Q_s\ll 1$ or $1/\left(BQ_s^2\right)\ll 1$.  (A line containing two quarks on opposite sides of the cut--corresponding to a gluon connecting a quark with its conjugate amplitude at a different coordinate--is not suppressed and therefore included in what follows.)  While these additional terms are formally higher order and will be ignored in the discussion to follow, we should point out that they were found to be important in obtaining a quantitative agreement with experiment~\cite{Dusling:2012cg}.

\begin{figure}[ht]
\scalebox{.4}{\includegraphics[]{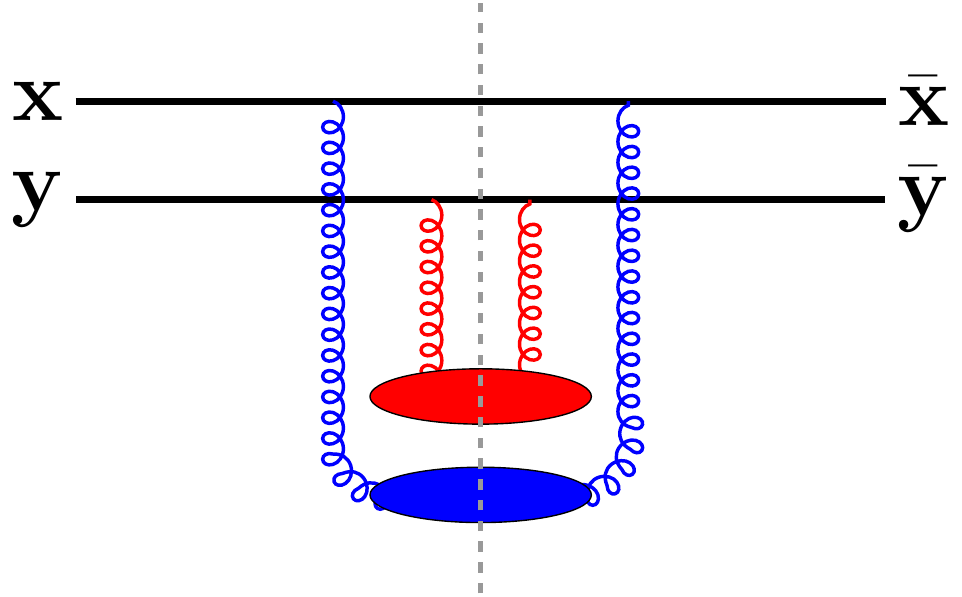}}
\scalebox{.4}{\includegraphics[]{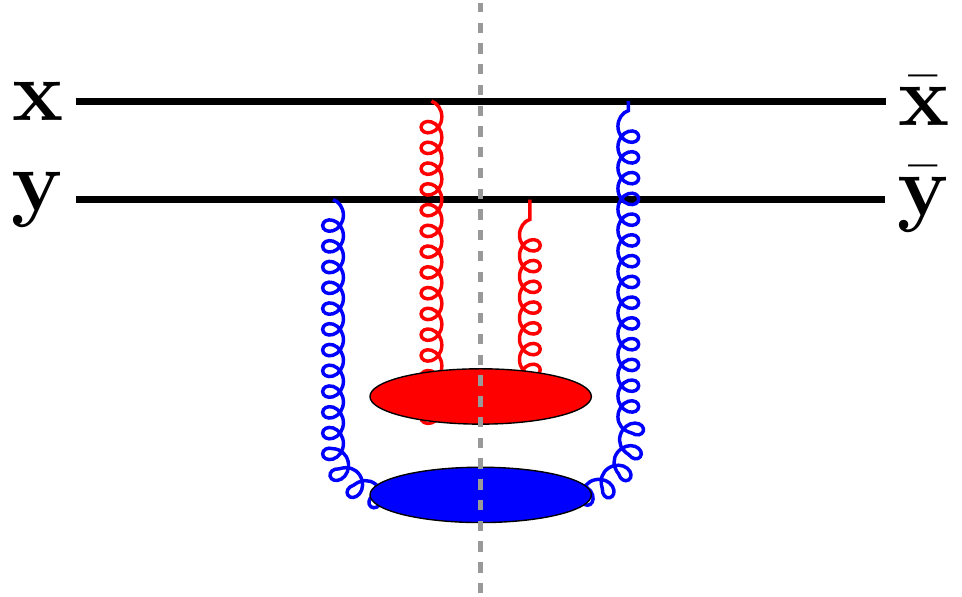}}
\scalebox{.4}{\includegraphics[]{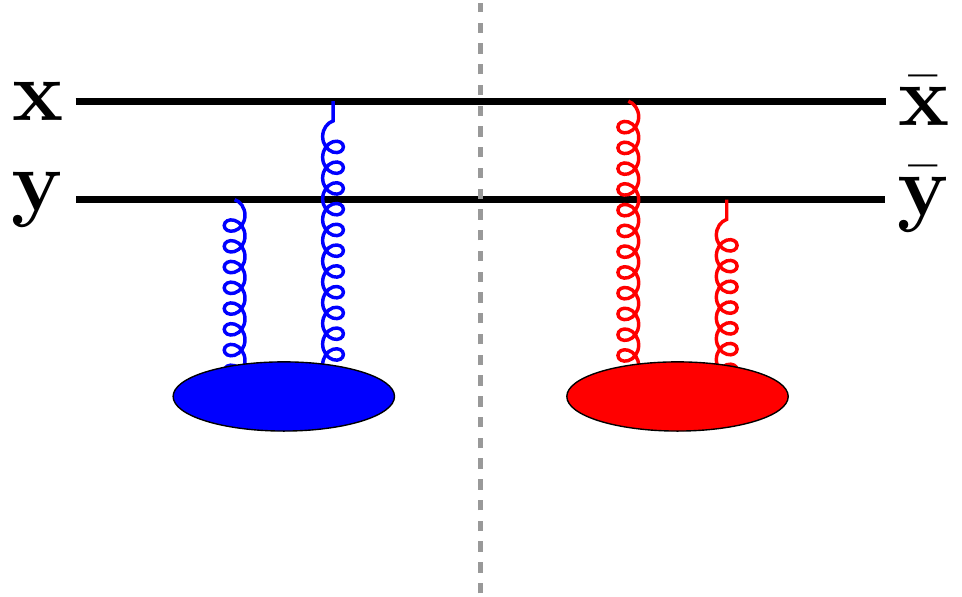}}
\caption{The three diagrams for two quarks scattering in the Glasma graph approximation.  The leftmost diagram is a disconnected contribution and equivalent to the square of single quark scattering. }
\label{fig:glasmagraphs}
\end{figure}

One can generate all possible Glasma graphs by starting with the completely disconnected diagram whereby each quark multiple scatters independently off the target.  We show an example of this completely disconnected contribution for two-quark scattering as the leftmost diagram of \Fig{fig:glasmagraphs}.

Further diagrams are generated by finding unique exchanges of coordinates (exchanging two gluon end points) along with an accompanying $\left(N_c^2-1\right)$ suppression for each exchange. There are two unique topologies for two quark scattering as shown in~\Fig{fig:glasmagraphs}.  The resulting expression for the Glasma graph approximation to the expectation value of two-dipole operators is therefore

\begin{eqnarray}
\left< D(\x,\xb)D(\y,\yb)\right> &=& D(\x,\xb)D(\y,\yb)+\frac{1}{\left(N_c^2-1\right)}\left[D(\xb-\y)D(\x-\yb)+D(\xb-\yb)D(\x-\y)\right]\,.
\label{eq:2partgg}
\end{eqnarray}

Higher-point correlators can be found in a similar fashion.  For $n$ quarks there are $(n-1)!!\equiv n(n-2)\cdots 3\cdot 1$ diagrams resulting from all possible unique contractions between quark lines.
For a correlation among six Wilson lines there are $5\cdot 3 \cdot 1 = 15$ pair-wise contractions while for
a correlator among eight Wilson lines there are $7\cdot 5\cdot 3\cdot1=105$ pair wise contractions.

As an aside, we remind the reader that the combinatorics discussed above are relevant for a dilute-dense framework.  A Glasma graph approximation has also been employed in the dense-dense limit which was shown to have $(n-1)!!^2$ diagrams for $n$-gluon production. In the dense-dense limit there would be $15^2=225$ and $105^2=11025$ diagrams for the six- and eight-point functions respectively as shown previously in Refs.~\cite{Dusling:2009ar} and \cite{Gelis:2009wh}.

\begin{figure}[ht]
\scalebox{.5}{\includegraphics[]{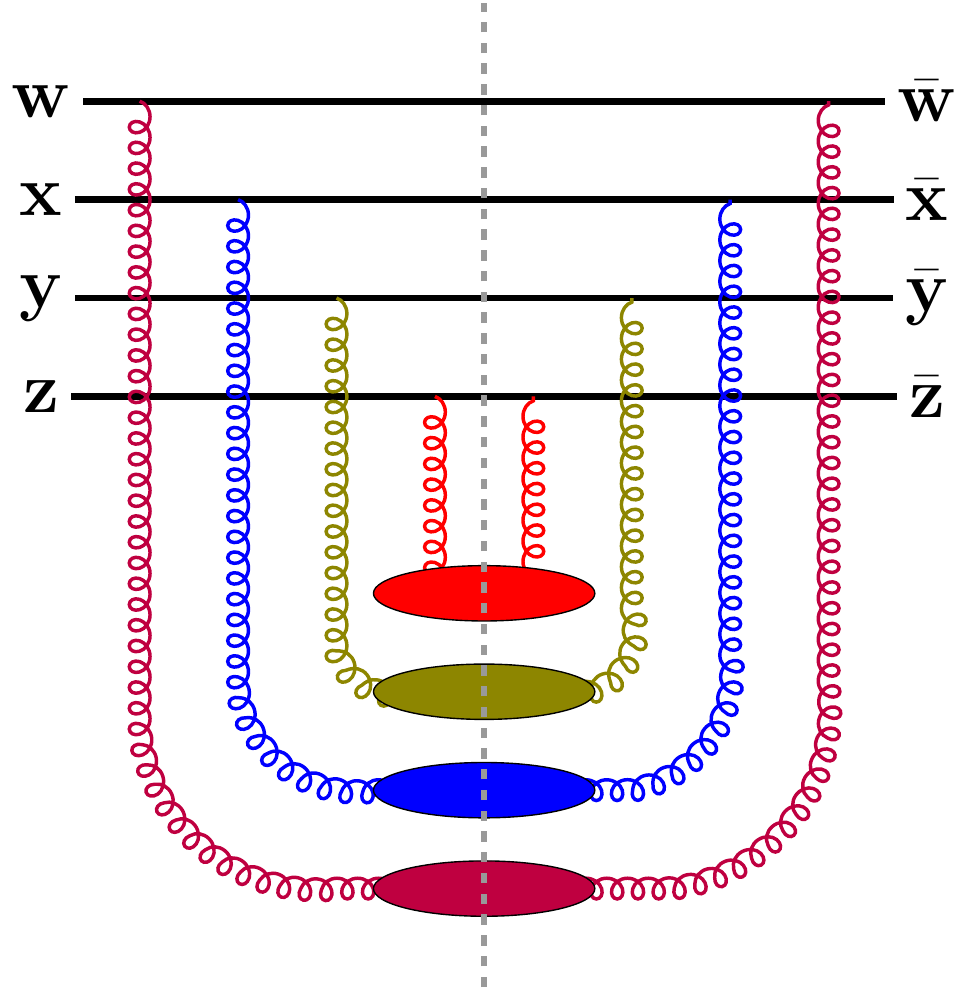}}\qquad\qquad
\scalebox{.5}{\includegraphics[]{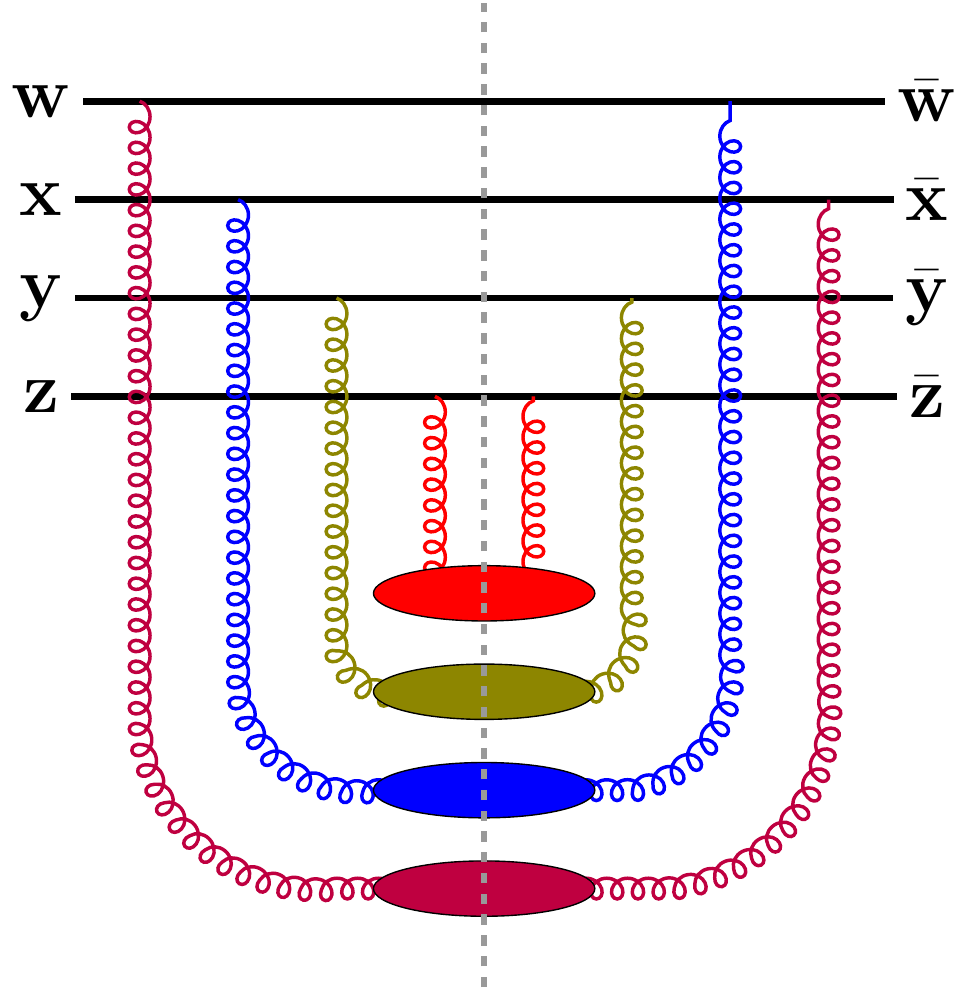}}
\caption{Left: Completely disconnected Glasma graph for four-quark scattering.  Right: Example of one of the 12 diagrams contained within $\mathcal{T}_1$ obtained from an exchange of one pair of coordinates--in this case $\xb$ and $\wb$.}
\label{fig:T1}
\end{figure}

We now come to the expectation value of four dipoles in the Glasma graph calculation following the procedure outlined above.  The starting point is the completely disconnected contribution in which each of the four quarks scatters independently as shown in the left diagram in \Fig{fig:T1}.

Starting from the disconnected diagram, there are 12 unique coordinate exchanges that can be made.  One example is shown in the right diagram of~\Fig{fig:T1} where the coordinates  $\xb$ and $\wb$ have been swapped.  These 12 diagrams resulting from a single coordinate exchange result in a contribution of $\left(N_c^2-1\right)^{-1} \mathcal{T}_1$ where
\begin{eqnarray}
\mathcal{T}_1=D(\w,\xb)D(\x,\wb)D(\y,\yb)D(\z,\zb)+\cdots \,.
\end{eqnarray}

\begin{figure}
\scalebox{.5}{\includegraphics[]{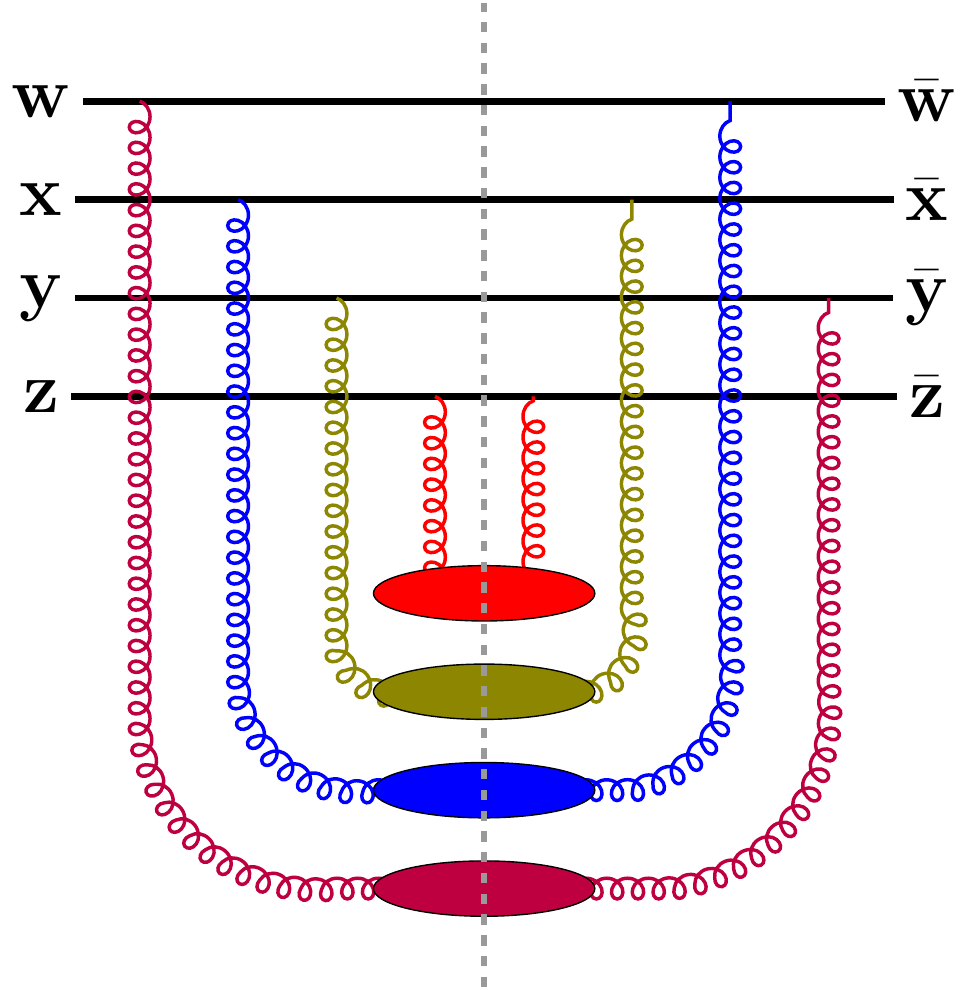}}\qquad\qquad
\scalebox{.5}{\includegraphics[]{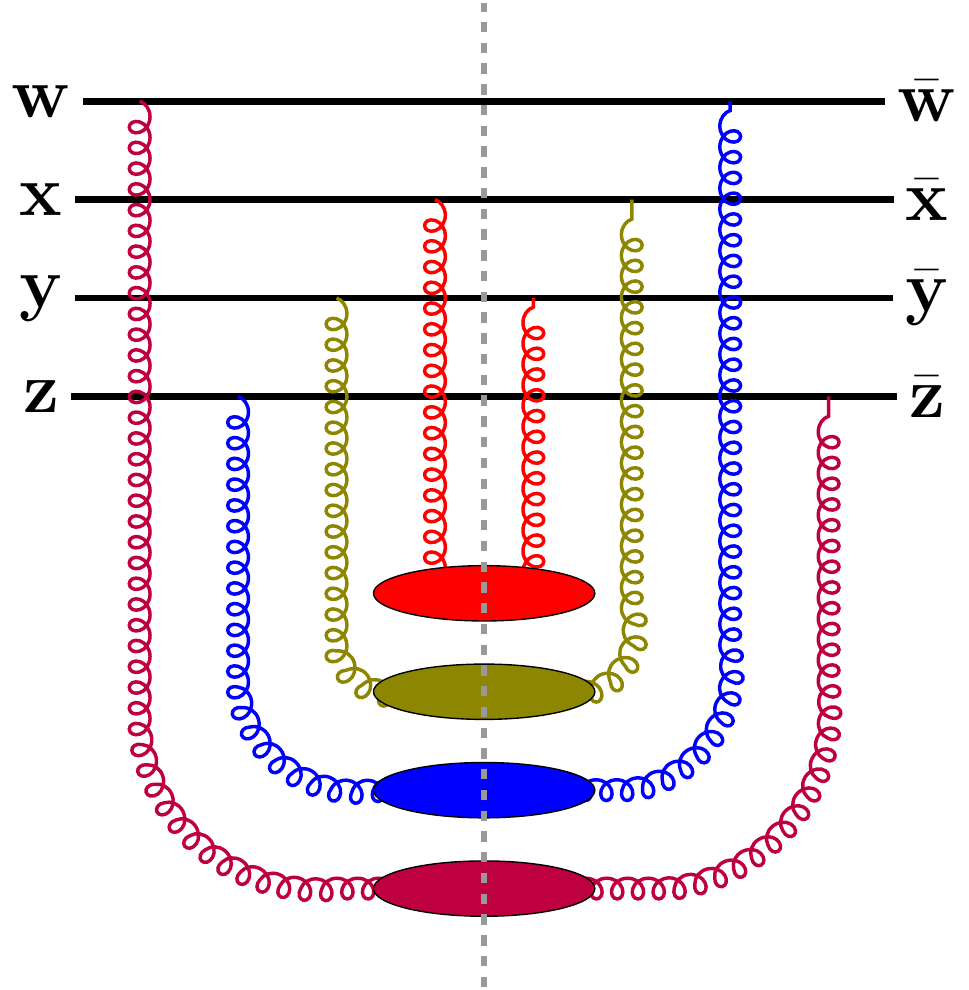}}
\caption{The diagram on the left is an example of one of 32 diagrams where three quarks are completely connected and one scatters independently--in this case the quark at $x$.  The diagram on the right shows one of 12 diagrams which factorizes into two two-dipole connected subgraphs.}
\label{fig:T2}
\end{figure}

We next consider diagrams resulting from unique and non-trivial two-coordinate exchanges.  There are two classes of diagrams which enter at this order.  The first is shown in the left diagram of \Fig{fig:T2}.  The diagram factorizes into two sub-graphs, one being a completely connected three-quark scattering and the other being a single independent quark scattering.  There are 32 such diagrams which will contribute with a factor of $\left(N_c^2-1\right)^{-2} \mathcal{T}_{2a}$ where
\begin{eqnarray}
\mathcal{T}_{2a}&=&D(\w,\yb)D(\x,\wb)D(\y,\xb)D(\z,\zb)+\cdots\,.
\end{eqnarray}

The second class of diagrams is shown in the right of~\Fig{fig:T2}.  It again factorizes into two sub-graphs, but in this case, each sub-graph is a completely connected two-quark scattering event ({\em i.e.} one of the two connected graphs shown in \Fig{fig:glasmagraphs}).  There are 12 such diagrams which will contribute as $\left(N_c^2-1\right)^{-2} \mathcal{T}_{2b}$ where
\begin{eqnarray}
\mathcal{T}_{2b}&=&D(\w,\zb)D(\z,\wb)D(\x,\yb)D(\y,\xb)+\cdots
\end{eqnarray}

\begin{figure}
\scalebox{.5}{\includegraphics[]{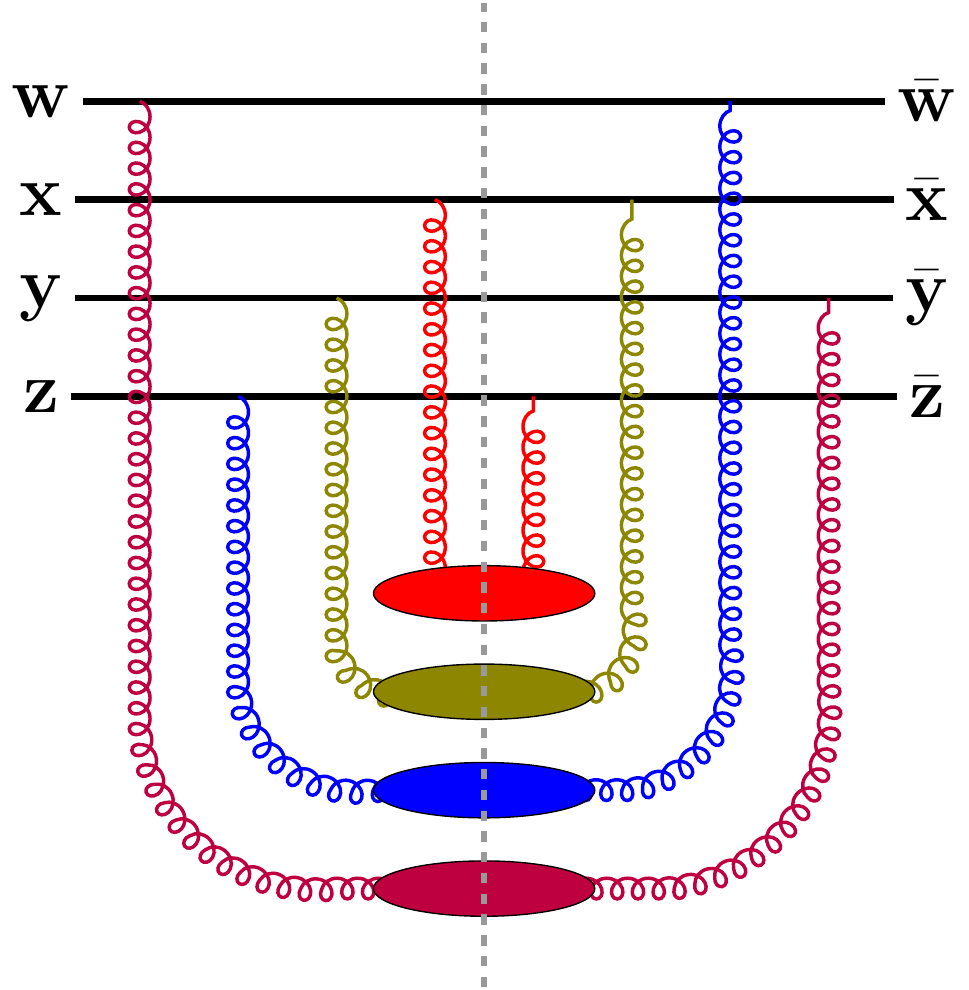}}
\caption{One of the 48 diagrams completely connected four-quark diagrams included in $\mathcal{T}_3$.}
\label{fig:T3}
\end{figure}

The last set of diagrams corresponds to those obtained by three unique and non-trivial coordinate exchanges.  An example of such a diagram is shown in \Fig{fig:T3}.  There are 48 such diagrams which contribute as  $\left(N_c^2-1\right)^{-3} \mathcal{T}_{3}$ where
\begin{eqnarray}
\mathcal{T}_3= D(\w,\yb)D(\z,\wb)D(\x,\zb)D(\y,\xb)+\cdots
\end{eqnarray}

Putting all of the diagrams together, we arrive at the final expression for the expectation value of four dipoles in the Glasma graph approximation,
\begin{eqnarray}
\left< D(\w,\wb)D(\x,\xb)D(\y,\yb)D(\z,\zb)\right> &=& D(\w,\wb)D(\x,\xb)D(\y,\yb)D(\z,\zb)+\frac{1}{\left(N_c^2-1\right)}\mathcal{T}_1(\w,\wb,\x,\xb,\y,\yb,\z,\zb)\nonumber\\*
&+&\frac{1}{\left(N_c^2-1\right)^2}\mathcal{T}_{2a}(\w,\wb,\x,\xb,\y,\yb,\z,\zb)+\frac{1}{\left(N_c^2-1\right)^2}\mathcal{T}_{2b}(\w,\wb,\x,\xb,\y,\yb,\z,\zb)\nonumber\\*
&+&\frac{1}{\left(N_c^2-1\right)^3}\mathcal{T}_3(\w,\wb,\x,\xb,\y,\yb,\z,\zb)\,.
\end{eqnarray}

We now come to the evaluation of the four-particle cumulant using the above Glasma graph approximation in the four-particle inclusive distribution, where  the four-particle cumulant is defined in~\Eq{eqn:cn4}. The Glasma graph approximation implicitly assumes that $\left(BQ_s^2\right)\gg 1$, and one must take this power counting into consideration when taking the ratios in the above expression for the cumulants.  In order to see this more clearly let us start with the expression for the double inclusive distribution,
\begin{eqnarray}
\frac{d^2N}{d^2\p{1}\;d^2\p{2}}=\frac{1}{(\pi \B)^2}\frac{1}{(2\pi)^4}
\int_{\rv{1}\rv{2}\R{1}\R{2}}
\left<D\left(\R{1}+\frac{r_1}{2},\R{1}-\frac{r_1}{2}\right)
D\left(\R{2}+\frac{r_2}{2},\R{2}-\frac{r_2}{2}\right)\right>\nonumber\\
\cdot\;e^{i\p{1}\cdot \rv{1}} e^{i\p{2}\cdot \rv{2}}
e^{-\sfrac{\rv{1}^2}{4\B}} e^{-\sfrac{\R{1}^2}{\B}}
e^{-\sfrac{\rv{2}^2}{4\B}} e^{-\sfrac{\R{2}^2}{\B}}\,.
\end{eqnarray}
We can evaluate the single inclusive multiplicity by integrating over $d^2\p{1}$ and $d^2\p{2}$ resulting in
\begin{eqnarray}
    N=\frac{1}{(\pi \B)^2}
    \int_{\R{1}\R{2}}
    \left<D\left(\R{1},\R{1}\right)
    D\left(\R{2},\R{2}\right)\right>
   e^{-\sfrac{\R{1}^2}{\B}}
   e^{-\sfrac{\R{2}^2}{\B}}\,.
\end{eqnarray}

In the full nonlinear theory (without approximation), we know that $\left<D\left(\R{1},\R{1}\right)D\left(\R{2},\R{2}\right)\right>=1$ and therefore $N=1$. However, in the Glasma graph approximation, the dipole-dipole correlator is replaced by \Eq{eq:2partgg} resulting in the following expression for the total multiplicity,
\begin{eqnarray}
    N&=&1+\frac{2}{(N_c^2-1)(\pi \B)^2}
    \int_{\R{1}\R{2}}D\left(\R{1}-\R{2}\right)^2
   e^{-\sfrac{\R{1}^2}{\B}}
   e^{-\sfrac{\R{2}^2}{\B}}\nonumber\\
   &=&1+\frac{2}{(N_c^2-1)}\frac{1}{\B Q^2+1}\,,
\end{eqnarray}
violating unitarity by a term of order $1/\left(BQ_s^2\right)\ll 1$.  When evaluating the cumulant in \Eq{eqn:cn4} one should formally only keep the leading order terms in $1/\left(BQ_s^2\right)$.  In practice, this means retaining only the leading disconnected contribution in the two terms $\kappa_0\{4\}$ and $\kappa_0\{2\}$ appearing in the denominators; these terms were defined in \Eq{eqn:kappa}.  For consistency within the Glasma graph approximation, we should therefore take $\kappa_0\{4\}= \left(\kappa_0\{2\}\right)^2= \left(\kappa_0\{1\}\right)^4$ where $\kappa_0\{1\}$ is just the single-inclusive distribution.

Now, coming to the numerator, due to rotational invariance, $\mathcal{T}_1$ and $\mathcal{T}_{2a}$ do not contribute to $\kappa_2\{4\}$.  Out of the 12 diagrams in $\mathcal{T}_{2b}$, four vanish by rotational invariance, and the remaining eight cancel with the term $2\left(\kappa_2\{2\}\right/\kappa_0\{2\})^2$.  So, within the Glasma graph approximation, the cumulant can be evaluated using
\begin{eqnarray}
c_2\{4\}=\frac{\kappa_2\{4\}\left[\mathcal{T}_3\right]}{\kappa_0\{1\}^4}\,,
\label{eq:c24gg}
\end{eqnarray}
where $\kappa_2\{4\}$ is evaluated using the 48 diagrams contained within $\mathcal{T}_3$.  With this, we compare this Glasma graph result with the full non-linear result introduced in Sec.~\ref{sec:dipolecorr} in~\Fig{fig:gg_sc} in the main text, with accompanying discussion in Sec.~\ref{sec:gg_comp}.

\end{document}